\documentclass{ieeetmlcn}
\usepackage{amsmath,amsfonts,amssymb}
\usepackage{algorithmic}
\usepackage{algorithm}
\usepackage{array}
\usepackage[caption=false,font=normalsize,labelfont=sf,textfont=sf]{subfig}
\usepackage{textcomp}
\usepackage{comment}
\usepackage{hyperref}
\usepackage{stfloats}
\usepackage{url}
\usepackage{verbatim}
\usepackage{graphicx,color}
\usepackage{cite}
\usepackage{xcolor}
\usepackage{tabularx}
\usepackage{booktabs}

\def\BibTeX{{\rm B\kern-.05em{\sc i\kern-.025em b}\kern-.08em
    T\kern-.1667em\lower.7ex\hbox{E}\kern-.125emX}}
\AtBeginDocument{\definecolor{tmlcncolor}{cmyk}{0.93,0.59,0.15,0.02}\definecolor{NavyBlue}{RGB}{0,86,125}}
\def\OJlogo{\vspace{-4pt}\includegraphics[height=18pt]{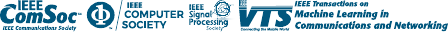}}
\def\seclogo{\vspace{10pt}\includegraphics[height=18pt]{new_logo}}
\def\authorrefmark#1{\ensuremath{^{\textbf{#1}}}}

\renewcommand{\thesection}{\Roman{section}}

\begin{document}
\receiveddate{XX Month, XXXX}
\reviseddate{XX Month, XXXX}
\accepteddate{XX Month, XXXX}
\publisheddate{XX Month, XXXX}
\currentdate{XX Month, XXXX}
\doiinfo{}
\markboth{}{M.Cheraghinia {et al.}}
\title{RFShape: An IQ-Pretrained Wireless Foundation Model for Heterogeneous IQ \& CIR Representations}
\author{Mohammad~Cheraghinia\authorrefmark{1},~\IEEEmembership{Graduate Student Member, IEEE}, Sebbe~Van~Der~Smissen\authorrefmark{2}, Jaron~Fontaine\authorrefmark{1}, Delia~Rico\authorrefmark{3}, Eli~De~Poorter\authorrefmark{1}, Adnan~Shahid\authorrefmark{1},~\IEEEmembership{Senior Member, IEEE}}
\affil{IDLab, Department of Information Technology, Ghent University - imec, 9052 Ghent, Belgium}
\affil{EDGX bv, Ghent, Belgium}
\affil{Universidad de Málaga, Andalucía Tech, 29071 Málaga, Spain}
\corresp{Corresponding author: Mohammad Cheraghinia (email: Mohammad.Cheraghinia@ugent.be).}
\authornote{This work was supported by the Horizon Europe program under the MCSA Staff Exchanges 2021 grant agreement 101086218 (EVOLVE project); the Horizon-JU-SNS-2023 Research and Innovation Program under Grant Agreement No. 101139194 (6G-XCEL project); the Fund for Scientific Research Flanders (FWO-Vlaanderen) under SB-PhD Fellowship with grant number 1S52025N and PESSO project with grant number G018522N; imec.icon project RAPIDNESS, which is co-financed by imec and Flanders Innovation and Entrepreneurship under project nr HBC.2024.0772; and imec.prospect program.}

\begin{abstract}
In wireless systems, foundation models have been proposed as alternative artificial intelligence models to replace task-specific models with a shared pretrained model that can be reused across many downstream applications (source separation, interference detection, localization, etc.). These wireless foundation models promise to reuse learned representations across diverse tasks, but choosing a foundation model also means choosing its input representation. For example, a model pretrained on spectrograms expects spectrograms, while one pretrained on in-phase/quadrature (IQ) samples expects IQ, limiting which pretrained model can be used for a given downstream problem. To remedy this, we present RFShape, a 7.4 M-parameter foundation model that breaks this coupling by adapting how a signal is presented to a shared pretrained encoder. RFShape is pretrained solely on synthetic IQ waveforms, without real captures, labels, or channel impulse responses (CIRs), and uses deterministic representation operators to support heterogeneous downstream inputs. Across ten tasks and nine datasets unseen during pretraining, the same encoder supports classification, ranging-error regression, and two-dimensional positioning. Representation selection alone increases 23-class radar recognition from 23.0\% to 69.5\% and short-range technology recognition from 70.8\% to 97.1\%. RFShape also outperforms CIR-pretrained ultra-wideband baselines by up to 12.7 percentage points for (N)LOS classification despite never observing CIRs during pretraining. These results show that foundation-model reuse need not be limited by the representation or modality seen during pretraining, and adapting how a signal is presented can extend a single pretrained encoder across heterogeneous wireless tasks.
\end{abstract}
\maketitle

\section{Introduction}
\label{sec:intro}
Wireless systems face a wide range of learning tasks, including recognizing which technology occupies a band, detecting interference, characterizing radar emissions, fingerprinting transmitters, and correcting localization errors. The standard practice is to train a dedicated model for each task and each input time scale. Every new task, band, radio, or sampling configuration therefore restarts data collection, architecture search, hyperparameter selection, and training, a process that is slow, costly, and prone to human error~\cite{cheraghinia2026wpfm}.

Foundation models lower this cost, but only for tasks whose inputs resemble the pretraining data. Existing wireless foundation models have begun to address heterogeneous and multimodal wireless inputs, but typically rely on including the supported representations or modalities during pretraining. This leaves open whether a model pretrained on a single representation can transfer to representations and physical modalities that were never observed during pretraining \cite{jiang2026largeAI}. A task whose natural input is a channel impulse response (CIR), or whose class-defining structure is visible only in the spectrum, therefore requires a new pretraining run, which removes the benefit the foundation model was meant to provide. 

This suggests an alternative to adapting the model: adapt how the signal is presented to it. Domain-informed input transformations, by contrast, require no training and complement the encoder’s learned features. Converting in-phase/quadrature (IQ) samples into a spectrum, an envelope, or an autocorrelation needs no training, no labels, and no hyperparameters. If a single encoder could accept all such views, the per-task effort would shift from training a model to choosing how the observation is presented to it.

We therefore ask how far a single frozen encoder can be carried by changing only what is presented to it. Two notions must be kept apart. 
We distinguish between a \textbf{modality}, the physical quantity reported by a receiver, such as an IQ waveform or CIR, and a \textbf{representation}, such as its spectrum, envelope, or autocorrelation. We hypothesise that, after normalisation and removal of scale and units, these representations share recurring local structures, such as bursts, decays, edges, and oscillations, which a codebook learned from IQ can capture across modalities. If the hypothesis holds, an encoder pretrained only on IQ waveforms should transfer across representations of the same modality and also to modalities it has never seen. In this work, the operator for each task is chosen from a small library of parameter-free transforms following a single principle to present the view in which the class-defining structure is stable. This is a search based on domain knowledge, but it is much cheaper than designing and training a new model.

To test this hypothesis, we present RFShape, a shape-codebook model whose 7.4\,M-parameter encoder is pretrained entirely on synthetic IQ waveforms including narrow-band Internet of Things (NB-IoT) uplink, fifth-generation (5G) non-terrestrial network (NTN) uplink, and 5G terrestrial downlink, all passed through a 3rd Generation Partnership Project (3GPP) NTN channel. Pretraining uses no real captures, no CIRs, and no labels. Because the encoder is univariate and applied independently to each channel, the number of channels is not an architectural parameter, and the same weights serve two-channel IQ, four-channel composite encodings, and 45-channel multi-anchor CIR stacks. We evaluate RFShape on ten downstream tasks over nine public and in-house datasets, grouped into three question families including \textbf{what kind of signal is this?} (wireless technology recognition (WTR), radar classification and characterization, modulation classification, interference detection, and transmitter fingerprinting), \textbf{how wrong is this measurement?} (ultra-wideband (UWB) ranging-error correction and line-of-sight/non-line-of-sight ((N)LOS) classification), and \textbf{where was this measured?} (position regression from multi-anchor CIRs). All downstream datasets are excluded from pretraining. Selected tasks evaluate representations absent from pretraining or transfer to the unseen CIR modality.

Our main contributions are as follows:
\begin{itemize}
\item \textbf{Representation-decoupled wireless foundation model.} We introduce RFShape, a wireless foundation model that decouples the pretrained encoder from the downstream input representation. Rather than adapting the backbone to each task, lightweight deterministic representation operators adapt how signals are presented to a shared encoder, enabling reuse across heterogeneous representations and modalities.
\item \textbf{Practical realization.} We demonstrate this concept with a single RFShape encoder pretrained exclusively on synthetic IQ waveforms, without real captures, CIRs, or task labels. The resulting model transfers to input representations absent from pretraining and to CIR measurements, a measurement modality excluded from pretraining, demonstrating flexibility in both input representation and measurement modality without changing the pretrained backbone.
\item \textbf{Performance evaluation.} We evaluate RFShape on ten tasks across nine public and in-house datasets, spanning classification, ranging-error regression, and two-dimensional positioning. All downstream datasets and task labels are excluded from pretraining, while selected settings assess transfer to unseen representations or the CIR modality.
\item \textbf{Discussion on representation and adaptation choices.} Finally, we systematically analyze input representations, latent readouts, frozen probes, feature fusion, and parameter-efficient adaptation to understand when and how representation-decoupled transfer succeeds. The results show that no single representation is optimal across all tasks, and that the benefit of additional adaptation varies across tasks, with several tasks addressed using a frozen encoder and lightweight probe.

\end{itemize}

The remainder of the paper is organized as follows. Section~\ref{sec:related} positions RFShape against existing wireless foundation models. Section~\ref{sec:ProbForm} formalizes the two-stage problem, the representation operator, and the pretraining objective. Section~\ref{sec:rfshape} describes the architecture, the pretraining corpus, and the four latent readouts. Section~\ref{sec:EXp} reports the ten downstream tasks, the representation and pretraining-channel ablations, and measured deployment cost. Finally, the conclusion summarizes the main findings.

\begin{table*}[!t]
\caption{Comparison with wireless foundation models discussed in
Section~\ref{sec:related}.}
\label{tab:related}
\centering
\footnotesize
\setlength{\tabcolsep}{2.5pt}
\renewcommand{\arraystretch}{1}
\begin{tabular}{@{}>{\raggedright\arraybackslash}p{3.2cm}
                  >{\centering\arraybackslash}p{1.5cm}
                  >{\raggedright\arraybackslash}p{2.7cm}
                  >{\raggedright\arraybackslash}p{5.6cm}
                  >{\centering\arraybackslash}p{1.5cm}
                  >{\centering\arraybackslash}p{2.5cm}@{}}
\toprule
Model & PT data & Pretrain input & Downstream input repr. & Cls. + Reg. & PT $=$ FT corpus \\
\midrule
\multicolumn{6}{@{}l}{\textbf{Pretraining on time-frequency images}}\\
MSM~\cite{aboulfotouh2024msm}
& Real & Spectrogram & Spectrogram (1) & yes$^{\dagger}$ & partly \\
WavesFM~\cite{aboulfotouh2025wavesfm}
& Real & Spectrogram, CSI & Spectrogram, CSI, OFDM grid (3) & yes & no \\
LWM-Spectro~\cite{kim2026lwmspectro}
& Synthetic & Spectrogram & Spectrogram (1) & no & no$^{\ddagger}$ \\
ACSF-TMAE~\cite{chen2025acsf}
& Mixed & Spectrogram & Spectrogram (1) & yes$^{\dagger}$ & yes$^{*}$ \\
\midrule
\multicolumn{6}{@{}l}{\textbf{Pretraining on raw IQ}}\\
SpectrumFM~\cite{zhou2026spectrumfm}
& Real & IQ & IQ (1) & no & yes$^{*}$ \\
IQFM~\cite{mashaal2025iqfm}
& Real & IQ & IQ (1) & no & yes$^{*}$ \\
WirelessJEPA~\cite{chu2026wjepa}
& Real & IQ & IQ (1) & no & yes$^{*}$ \\
Multimodal WFM~\cite{aboulfotouh2026multimodal}
& Real & Spectrogram, IQ & Spectrogram, CSI, IQ (3) & yes & partly \\
EMind~\cite{emind2026}
& Mixed & IQ & IQ (1) & yes & yes$^{*}$ \\
WTR/loc.\ FM~\cite{cheraghinia2025}
& Real & IQ, CIR & IQ, CIR (2) & yes & yes$^{*}$ \\
Lightweight MLP FM~\cite{cheraghinia2025lwfm}
& Real & IQ, CIR & IQ, CIR (2) & no & yes$^{*}$ \\
\midrule
\multicolumn{6}{@{}l}{\textbf{Language-model interfaces to RF}}\\
RadioLLM~\cite{chen2025radiollm}
& Mixed & IQ & IQ (1) & yes$^{\dagger}$ & yes$^{*}$ \\
RFLM~\cite{wen2026rflm}
& Real & Multi-view IQ & IQ, spectrogram, constellation (3) & yes$^{\dagger}$ & yes$^{*}$ \\
RF-GPT~\cite{zou2026rfgpt}
& Synthetic & Spectrogram & Spectrogram (1) & no & yes$^{*}$ \\
Seeing Radio~\cite{zou2026seeing}
& Synthetic & Spectrogram, IQ plot & Spectrogram, IQ plot (2) & no & yes$^{*}$ \\
PReD~\cite{han2026pred}
& Mixed & Multi-view images & IQ plot, FFT, STFT, constellation (4) & yes$^{\dagger}$ & yes$^{*}$ \\
\midrule
\textbf{RFShape (ours)}
& \textbf{Synthetic}
& \textbf{IQ}
& \textbf{IQ, CIR, FFT, ACF, composite radar, \ldots\ (7+)}
& \textbf{yes} & \textbf{no} \\
\bottomrule
\end{tabular}
\vspace{2pt}
{\raggedright \scriptsize
PT $=$ pretraining, FT $=$ fine-tuning. \\
$^{*}$Pretraining and fine-tuning using same dataset; generalization is tested by holding out a class or a capture location rather than a new domain. \\
$^{\dagger}$The non-classification tasks are signal reconstruction (denoising, recovery, source separation) or text-generated estimates, not trained scalar regression. \\
$^{\ddagger}$Evaluation uses a held-out city scenario, but the pretraining and
evaluation signals come from the same generator and the same spectrogram
representation.\par}
\end{table*}

\section{Related Work}
\label{sec:related}
Wireless foundation models differ mainly in the input they are pretrained on, because that choice decides which downstream inputs the model can later accept. We organize related works along this axis.
\subsection{Pretraining on time-frequency images}
Several models adapt vision architectures to process signals as time-frequency images. For example, masked spectrogram modeling (MSM)~\cite{aboulfotouh2024msm} and WavesFM~\cite{aboulfotouh2025wavesfm} apply masked modeling to spectrograms and orthogonal frequency-division multiplexing (OFDM) resource grids using convolutional long short-term memory networks (ConvLSTMs) and Vision Transformers, respectively, to address diverse tasks including spectrum forecasting, 5G positioning, and classification. Two further models pretrain transformer autoencoders on spectrograms but differ in both objective and scope. LWM-Spectro~\cite{kim2026lwmspectro} combines masked modeling with a contrastive loss in a mixture-of-experts backbone, and targets modulation classification and joint signal-to-noise ratio (SNR) and mobility recognition. ACSF-TMAE~\cite{chen2025acsf} instead adds mutual-information maximization and an attention-based channel-spectrum fusion to a masked autoencoder, which lets it extend beyond classification to denoising and recovery of missing signal segments. Crucially, in \textbf{all} these approaches the time-frequency transform is fixed before pretraining, so a downstream task must present its data as the same kind of image, even where several image-like modalities are supported.
\subsection{Pretraining on raw IQ}
A second group avoids fixed transforms by pretraining directly on IQ samples. For example, SpectrumFM~\cite{zhou2026spectrumfm} pretrains with masked reconstruction and next-slot prediction, then adapts to modulation classification, wireless technology classification, spectrum sensing, and anomaly detection. EMind~\cite{emind2026} instead pursues scale, pretraining on 81 million samples drawn from 14 sources to support modulation classification, parameter regression, radio-frequency (RF) fingerprinting, and interference recognition. On the other hand, IQFM~\cite{mashaal2025iqfm} applies contrastive learning with task-aware augmentations for few-shot transfer, whereas WirelessJEPA~\cite{chu2026wjepa} simplifies this using augmentation-free latent prediction over an antenna--time grid. Recent extensions also improve deployment efficiency via early-exit task heads~\cite{mashaal2026earlyexit} and multimodal IQ-image encoders~\cite{aboulfotouh2026multimodal}. Closest to our present work, our earlier models pretrain on unlabeled IQ and CIR data for technology recognition and ranging~\cite{cheraghinia2025}, subsequently adapting transformers into lightweight patch-wise multilayer perceptrons (MLPs) for edge deployment~\cite{cheraghinia2025lwfm}. However, across this entire group, downstream inputs are restricted to the modalities seen during pretraining, and the pretraining dataset typically overlaps with the fine-tuning data.
\subsection{Language-model interfaces to RF}
A third, recent group does RF processing through large language models (LLMs). Studies like RadioLLM~\cite{chen2025radiollm} and RFLM~\cite{wen2026rflm} map raw IQ signals directly into the token space of instruction-following LLMs for tasks ranging from classification to blind source separation. Alternatively, RF-GPT~\cite{zou2026rfgpt},~\cite{zou2026seeing}, and PReD~\cite{han2026pred} leverage the visual encoders of multimodal LLMs by rendering IQ data as diverse plots (e.g., spectrograms, constellations). Two crucial insights from these works inform our approach. First, different views of the same signal expose different task-relevant information; second, fully synthetic datasets can successfully train general RF models, as also demonstrated by generative approaches like RF-Diffusion~\cite{chi2024rfdiffusion}. However, relying on massive LLM backbones incurs severe inference latency and memory costs, rendering them unsuitable for processing near the air interface~\cite{navidan2026oran}.
\subsection{Positioning of RFShape}
Table~\ref{tab:related} positions RFShape against recent wireless foundation models based on input representation diversity, support for regression, and dataset overlap between pretraining and fine-tuning. This highlights three critical gaps. First, prior models evaluate on downstream inputs from their pretraining modality, whereas we evaluate a fixed model across completely different unseen representations including FFT, autocorrelation, composite radar encodings, and UWB CIRs. Second, about half are classification-only, and several of the rest reach
beyond classification only through signal reconstruction or text-generated
estimates rather than trained scalar regression, while several of our
tasks are genuine regressions, including ranging error, pulse parameters, and position. Third, generalization should be distinguished along three axes including transfer to new datasets, transfer to input representations absent from pretraining, and transfer to a measurement modality absent from pretraining. Even the few that generalize to new datasets (e.g., WavesFM, LWM-Spectro) remain constrained to their original input modality. In contrast, RFShape strictly holds out \textbf{both} the domain and the input representation. Section~\ref{sec:wtr} quantifies the cost of this generalization regime.

\section{Problem Formulation}
\label{sec:ProbForm}
\subsection{System model}
\begin{figure}
    \centering
    \includegraphics[width=\linewidth]{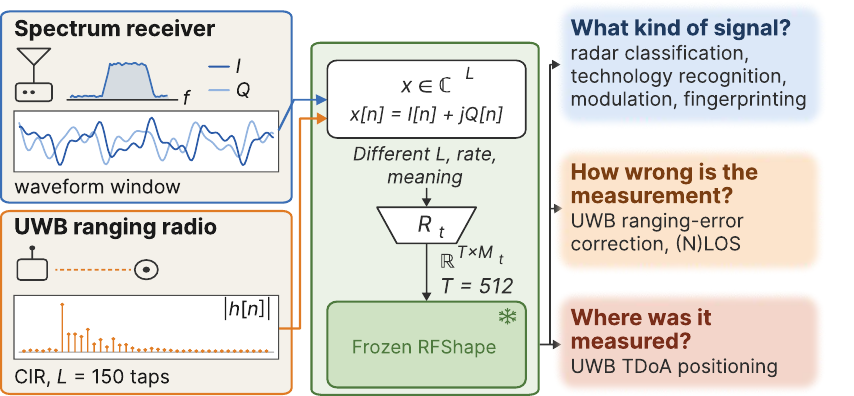}
    \caption{Different receivers output complex sequences $x \in \mathbb{C}^{L}$ of varying length and meaning, which $\mathcal{R}_t$ maps to a common $T = 512$ representation for a frozen RFShape backbone serving multiple downstream tasks.}
    \label{fig:system_model}
\end{figure}
Fig.~\ref{fig:system_model} illustrates the system model. A receiver outputs a finite complex baseband sequence
$x = [x[0], \dots, x[L-1]] \in \mathbb{C}^{L}$, $x[n] = I[n] + jQ[n]$. Two physically different receivers in this paper share that description including a waveform window sampled from a live band, and a CIR reported by a ranging radio. They differ in length ($L$), sample rate and meaning, but not in type. Across the tasks studied, $L$ ranges from a 150-tap CIR to a 4096-sample synthetic NTN window. UWB devices report a fixed accumulator window, 150 and 300 taps in datasets are not a part of the modelling choice and the informative part in UWB CIR is small due to short range of this technology. At the long end the pretraining window is a design parameter, set to 4096 complex samples. Integrating that heterogeneity is therefore the first problem to solve, and it motivates a per-task representation operator $\mathcal{R}_t$ that brings every observation to the common length and scale.

A \textbf{task} is a pair $t = (\mathcal{D}_t, \mathcal{Y}_t)$ consisting of a labeled dataset $\mathcal{D}_t = \{(x_i, y_i)\}$ and a target space $\mathcal{Y}_t$, where each $x_i$ is an input and $y_i \in \mathcal{Y}_t$ is its corresponding label. The target space can be categorical $\mathcal{Y}_t = \{1, \dots, C_t\}$, continuous $\mathcal{Y}_t = \mathbb{R}^{d_t}$, or a product of the two. We study 10 tasks over nine datasets (details are in Table~\ref{tab:tasks}) and Table~\ref{tab:notation} collects the symbols and the value each takes in this work.
\subsection{Problem statement}
The conventional response to a set of tasks $\mathcal{T}$ is $|\mathcal{T}|$ models. Each is trained from scratch on its own $\mathcal{D}_t$, which multiplies data collection, architecture search and deployment by $|\mathcal{T}|$. We ask instead what a single encoder can do if it is trained once and then never touched.
Let $\mathcal{D}_0$ be an unlabelled dataset and let $f_\theta$ be a univariate
encoder. Stage one solves
\begin{equation}
\theta^{\star} = \arg\min_{\theta}\;
\mathcal{L}_{\mathrm{pre}}(\theta;\, \mathcal{D}_0),
\label{eq:stage1}
\end{equation}
where $\mathcal{L}_{\mathrm{pre}}$ is the self-supervised objective defined in~\eqref{eq:pretrain}. Stage two then solves, separately for each task,
\begin{equation}
\, \psi_t^{\star} \;=\;
\arg\min_{\psi_t}\;
\mathcal{L}_t\big(\mathcal{R}_t,\, \psi_t;\; \mathcal{D}_t,\, \theta^{\star}\big),
\qquad t \in \mathcal{T},
\label{eq:stage2}
\end{equation}
where $\mathcal{L}_t$ is the loss of task $t$ and $\mathcal{D}_t$ its labeled dataset. The pretrained parameters $\theta^{\star}$ are held fixed and are not optimized during this stage; their position to the right of the semicolon marks them as a parameter of the objective rather than an optimization variable. It is the same $\theta^{\star}$ in all $|\mathcal{T}|$ problems, so the only thing that distinguishes one task from another is the pair $ \psi_t$ that is learned by optimization.

\begin{table}[!t]
\caption{Notation and the value each quantity takes in this work.
\label{tab:notation}}
\centering
\footnotesize
\setlength{\tabcolsep}{4pt}
\begin{tabularx}{\columnwidth}{@{}l X l@{}}
\toprule
Symbol & Meaning & Value \\
\midrule
$x \in \mathbb{C}^{L}$ & complex observation window & $L \in [150, 4096]$ \\
$\mathcal{T}$ & set of downstream tasks & $|\mathcal{T}| = 10$ \\
$C_t$ & classes, task $t$ & 2--37 \\
$d_t$ & regression targets, task $t$ & 1--4 \\
$\mathcal{D}_0$ & pretraining dataset & synthetic NTN IQ \\
$\rho$ & masked-patch fraction & 0.185 \\
$\mathcal{R}_t$ & representation operator & Table~\ref{tab:ops} \\
$M_t$ & real channels after $\mathcal{R}_t$ & 2--45 \\
$T$ & model input length & 512 \\
$L_p$, $P$ & patch length, patches per window & 8, 64 \\
$f_\theta$ & frozen encoder + tokenizer & 7.4\,M param. \\
$N$ & shape tokens per channel & 32 \\
$K, d_c$ & codebook size, code dimension & 1024, 8 \\
$D$ & embedding dimension & 256 \\
$g$ & per-channel latent descriptor, width $d_g$ & \eqref{eq:readouts} \\
$\phi(x)$ & concatenated task feature, $\mathbb{R}^{M_t d_g}$ & 256--46080 \\
$h_\psi$ & task head & linear / 1-layer MLP \\
$r, \alpha$ & LoRA rank, scaling & 16, 4 \\
\midrule
\multicolumn{3}{@{}p{\columnwidth}@{}}{\footnotesize LoRA denotes low-rank adaptation. Index convention: $t$ subscripts task-level objects ($\mathcal{R}_t$, $M_t$, $C_t$); $i$ indexes the $N$ shape tokens; $m$ indexes the $M_t$ channels; $k$ indexes the $K$ codes.} \\
\bottomrule
\end{tabularx}
\end{table}

What remains to be determined is how an observation is presented to the encoder, using a representation operator \(\mathcal{R}_t\) (see Sec.~\ref{sec:repops}). Additionally, a small task head \(h_{\psi_t}\) is utilized to convert the encoder's output into the target, where the size of \(|\psi_t|\) is much smaller than \(|\theta^{\star}|\) (refer to Sec.~\ref{sec:adapt}).

\textbf{The pretraining dataset is disjoint from every task.} We require $\mathcal{D}_0 \cap \mathcal{D}_t = \emptyset$ for all $t$, and more strongly that no downstream data appears in pretraining dataset $\mathcal{D}_0$ which is fully synthetic, IQ-only, and passed through a single NTN channel, so the model has seen no real capture, no CIR, and no label of any kind at the time stage two is solved. This setup evaluates transfer from synthetic pretraining data to separate downstream datasets without exposing the encoder to downstream observations during pretraining.

\textbf{The encoder is frozen.} $\theta^{\star}$ is trained once, in stage one, and is never updated for a task. Only the two variables of~\eqref{eq:stage2} change from one task to the next, and the same weights serve every task in $\mathcal{T}$. We later admit one bounded relaxation of this constraint: a low-rank correction $W' = W + \Delta W$ to selected projections of $\theta^{\star}$, which leaves the pretrained weights $W$ themselves unchanged (Sec.~\ref{sec:adapt}, \eqref{eq:lora}).

Solving stage two under those two properties raises two concrete difficulties, and the rest of the paper is organized around them:

\textbf{(P1) Input heterogeneity.} Tasks differ in modality (IQ or CIR), length ($L$ from 150 to 4096), sample rate, and channel count ($M_t$ from 2, where I and Q are two real series, to 45 for the multi-anchor time difference of arrival (TDoA) stack of 15 anchor slots $\times$ 3 maps), while $f_\theta$ is one univariate model expecting a fixed length $T$. The operator $\mathcal{R}_t$ and per-channel encoding absorb all of this without touching $\theta$.

\textbf{(P2) Target heterogeneity.} A single task may require a class label and several real-valued quantities in different physical units. The head and the loss weighting should handle this.

\subsection{Representation operators}
\label{sec:repops}
The operator $\mathcal{R}_t: \mathbb{C}^{L} \rightarrow \mathbb{R}^{T \times M_t}$ stacks $M_t$ scalar maps, resamples each to the model input length $T$, and applies a per-channel $z$-score,
\begin{equation}
\mathcal{R}_t(x) \;=\; \mathrm{z}\Big(\mathrm{resample}_T
\big[\, r_1(x) \;\; \cdots \;\; r_{M_t}(x) \,\big]\Big).
\label{eq:repop}
\end{equation}
where $r_m: \mathbb{C}^{L} \rightarrow \mathbb{R}^{L}$ is the $m$-th scalar map
of Table~\ref{tab:ops} and $\mathrm{resample}_T$ brings each map to the model
length $T$ (linear or anti-aliased polyphase), and
$\mathrm{z}(\cdot)$ subtracts the mean and divides by the standard deviation of
each channel separately, so channels of very different dynamic range reach the encoder on a common scale. Table~\ref{tab:ops} lists the maps used in this paper. A task selects a subset and concatenates it. The discrete Fourier transform (DFT) is denoted by $\mathrm{DFT}$. The IQAP encoding stacks the in-phase, quadrature, amplitude, and differential-phase channels, $[r_{\mathrm{I}}, r_{\mathrm{Q}}, r_{\mathrm{A}}, r_{\mathrm{DP}}]$; its P channel denotes differential phase rather than absolute phase. As an example, the RADAR composite is $[r_{\mathrm{LOGA}}, r_{\mathrm{CIF}}, r_{\mathrm{ACF}}, r_{\mathrm{SPECC}}]$ with $M_t{=}4$, the UWB tasks use $[r_{\mathrm{I}}, r_{\mathrm{Q}}, r_{\mathrm{A}}]$ on the CIR, and the short-range WTR view applies $[r_{\mathrm{I}}, r_{\mathrm{Q}}, r_{\mathrm{A}}]$ to the spectrum $X = \mathrm{DFT}(x)$ rather than to $x$.

\begin{table}[!t]
\caption{Scalar maps $r_m$ applied to a complex sequence
$z$. $\tilde{z}[n] = z[n]e^{-j\hat{\omega}n}$ is $z$ with its estimated
carrier removed, $\hat{\omega} = \arg\sum_n z[n{+}1]\overline{z[n]}$, and
$R_{zz}[\tau] = \sum_n z[n{+}\tau]\overline{z[n]}$ is its
autocorrelation.\label{tab:ops}}
\centering
\footnotesize
\setlength{\tabcolsep}{3pt}
\begin{tabular}{@{}
  >{\raggedright\arraybackslash}p{1.10cm}
  >{\raggedright\arraybackslash}p{2.6cm}
  >{\raggedright\arraybackslash}p{2.8cm}
  >{\centering\arraybackslash}p{1.5cm}
  @{}}
\toprule
Map & Name & Definition & Tasks \\
\midrule
$r_{\mathrm{I}}$    & In-phase & $\Re\, z[n]$ & 1--7, 9, 10  \\
$r_{\mathrm{Q}}$    & Quadrature & $\Im\, z[n]$ & 1--7, 9, 10  \\
$r_{\mathrm{A}}$    & Envelope & $|z[n]|$ & 1, 4--7, 9, 10  \\
$r_{\mathrm{LOGA}}$ & Log-envelope & $\log\!\big(1 + |z[n]| / \mathrm{rms}(z)\big)$ & 2  \\
$r_{\mathrm{ACF}}$  & Autocorrelation & $|\rho[\tau]| \,/\, |\rho[0]|$ & 2, 8 \\
$r_{\mathrm{DP}}$   & Differential phase & $\arg\!\big(z[n]\,\overline{z[n{-}1]}\big)$ & 1, 7  \\
$r_{\mathrm{CIF}}$  & De-rotated frequency & $\arg\!\big(\tilde{z}[n]\,\overline{\tilde{z}[n{-}1]}\big)$ & 2\\
$r_{\mathrm{SPEC}}$ & Log-spectrum & $\log\!\big(1 + |\mathrm{DFT}(z)[k]|\big)$ & 8 \\
$r_{\mathrm{QA}}$   & Envelope quantiles & $\mathrm{sort}\big(|z[n]|\big)$ & 8 \\
$r_{\mathrm{QI}}$   & In-phase quantiles & $\mathrm{sort}\big(\Re\, z[n]\big)$ & 8 \\
$r_{\mathrm{SPECC}}$& De-rotated spectrum & $\log\!\big(1 + |\mathrm{DFT}(\tilde{z})[k]|\big)$ & 2 \\
\bottomrule
\end{tabular}
\end{table}

\begin{figure*}
    \centering
    \includegraphics[width=\linewidth]{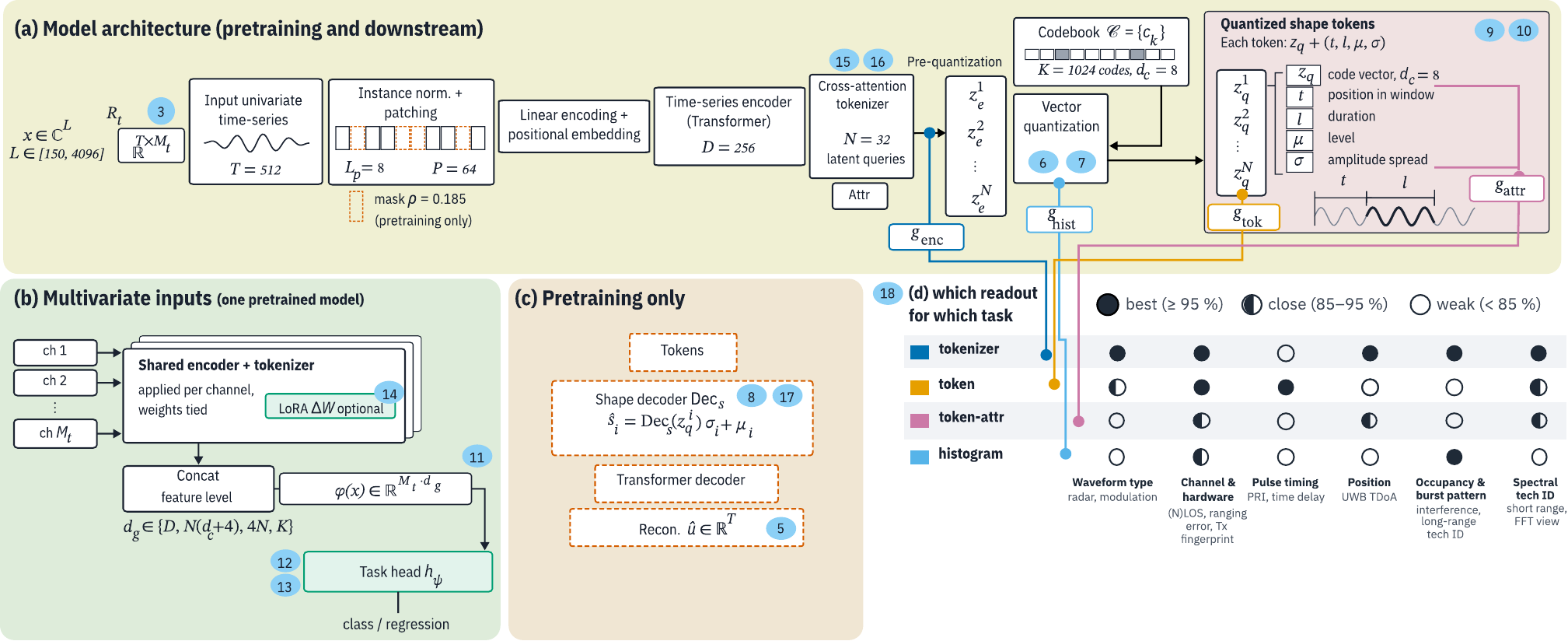}
    \caption{Overview of the RFShape architecture. Blue circles denote the corresponding equation numbers for each component. (a) A univariate RFShape that is used in both pretraining and downstream tasks. (b) Multivariate inputs are handled by applying the same encoder and tokenizer per channel and concatenating at the feature level. (c) The Transformer decoder is used only to compute the reconstruction term of~\eqref{eq:pretrain} and is discarded in downstream. (d) The readouts and their suitability for different types of downstream tasks.}
    \label{fig:arch}
\end{figure*}

\subsection{Pretraining objective}
Given unlabeled windows $\{x_i\}$, each channel $u \in \mathbb{R}^{T}$ is normalized, a random fraction $\rho$ of its patches is masked. The model $f_\theta$ must reconstruct the series using an encoder that produces embeddings, then a cross-attention tokenizer emits $N$ latent tokens. Each token is vector-quantized against a codebook $\mathcal{C}=\{c_k\}_{k=1}^{K}$ together with explicit attributes $(t_i, l_i, \mu_i, \sigma_i)$ that give the token's normalized position, length, offset, and scale. This pipeline is depicted in Fig.~\ref{fig:arch}(a), with the reconstruction branch that closes the objective shown in Fig.~\ref{fig:arch}(c). Training minimizes
\begin{equation}
\mathcal{L}_{\mathrm{pre}} \;=\; \mathcal{L}_{\mathrm{rec}}
 + \lambda_1 \mathcal{L}_{\mathrm{vq}}
 + \lambda_2 \mathcal{L}_{\mathrm{shape}}
 + \lambda_3 \mathcal{L}_{\mathrm{dist}},
\label{eq:pretrain}
\end{equation}
\noindent where $\lambda_1, \lambda_2, \lambda_3$ are fixed scalar weights
(Table~\ref{tab:hyper}). The four loss terms jointly encourage accurate signal reconstruction, discrete shape encoding, local shape matching, and diverse temporal coverage, as detailed below:
\subsubsection{Reconstruction} 
The reconstruction loss measures how accurately the model recovers the original, masked channel. 
\begin{equation}
    \mathcal{L}_{\mathrm{rec}} = \lVert \hat{u} - u
\rVert_2^2 / T,
\label{eq:reconloss}
\end{equation}
where $\hat{u} \in \mathbb{R}^{T}$ is the decoder's estimate of
the unmasked channel $u$. This is the only term that uses the decoder of
Fig.~\ref{fig:arch}(c).
\subsubsection{Quantization} The quantization bottleneck maps the continuous representations generated by the encoder into a discrete latent space using a shared, learned codebook. Let $z_e^i \in \mathbb{R}^{d_c}$ denote the $i$-th continuous pre-quantization token output by the encoder, where $d_c$ is the feature dimension. For each encoded token, we compute its Euclidean distance to all available embedding vectors $c_k$ in the codebook to find the nearest neighbor. The index of the matching entry is identified as $k_i = \arg\min_k \lVert z_e^i - c_k \rVert_2$. We then discretize by replacing the continuous token with this matched codebook vector. The final quantized token is $z_q^i = c_{k_i}$. The quantized shape token is then the matched code together with its attributes, $(z_q^i, t_i, l_i, \mu_i, \sigma_i) \in \mathbb{R}^{d_c+4}$, the object labeled \emph{quantized shape tokens} in Fig.~\ref{fig:arch}(a). The set of $N$ such tokens is what the decoder reconstructs from during pretraining, and what the descriptor $g_{\mathrm{tok}}$ exposes downstream. 

To effectively train this non-differentiable bottleneck and align the encoder outputs with the codebook embeddings, the vector quantization loss is defined as:
\begin{equation}
\mathcal{L}_{\mathrm{vq}} =
\big\lVert \mathrm{sg}[z_q] - z_e \big\rVert_2^2
+ \beta \big\lVert z_q - \mathrm{sg}[z_e] \big\rVert_2^2
+ \eta\,\mathcal{H},
\label{eq:vq}
\end{equation}
where $\mathrm{sg}[\cdot]$ is the stop-gradient operator, the first term pulls
the encoder towards the codebook, and the second pulls the codebook towards the
encoder, and $\beta$ balances the two. Note that $\beta$ weights the codebook term here. Because the minimization process has no useful
gradient, the encoder is updated through the straight-through estimator
$z_q \leftarrow z_e + \mathrm{sg}[z_q - z_e]$, which passes the decoder's
gradient to $z_e$ unchanged. The last term keeps the codebook usable, as defined below:
\begin{multline}
p^i_k = \mathrm{softmin}_k\!\big(\lVert z_e^i - c_k\rVert_2^2 / \kappa\big),
\qquad \\
\mathcal{H} = \mathbb{E}_i\big[\mathbb{H}(p^i)\big]
 - \gamma\, \mathbb{H}\big(\mathbb{E}_i[p^i]\big),
\label{eq:entropy}
\end{multline}
$p^i$ is a distribution over the $K$ codes and $\mathbb{H}(\cdot)$ is the Shannon entropy. Minimizing the first entropy makes each token commit to a single code and maximizing the entropy of the batch marginal stops the model from collapsing. This is what makes the usage histogram informative rather than degenerate. The $\beta, \eta, \gamma$ are the codebook-update, entropy, and marginal weights and $\kappa$ is soft-assignment temperature.
\subsubsection{Shape matching} The shape matching objective ensures that each token accurately captures and reconstructs its corresponding segment of the input signal. Conceptually, each token corresponds to a specific temporal window of the input $u$. To formalize this, let $\Pi_{t_i, l_i}(u) \in \mathbb{R}^{L_s}$ denote the sub-series of the input extracted over the normalized interval $[t_i, t_i{+}l_i]$. To enable uniform processing and shape comparison, we resample this segment to a fixed sequence length $L_s$. 

We define this standardized ground-truth segment as $s_i$, and compute its empirical mean $\mu_i^{\star}$ and standard deviation $\sigma_i^{\star}$. These calculated local statistics serve as the ground-truth to supervise the token's predicted shift and scale. Finally, let $\bar{s}_i = \mathrm{Dec}_s(z_q^i)$ denote the normalized shape profile that is decoded directly from the quantized token, before it is given an offset and a scale,
\begin{equation}
\mathcal{L}_{\mathrm{shape}} =
\Big\lVert \bar{s}_i - \mathrm{sg}\big[\tfrac{s_i - \mu_i^{\star}}{\sigma_i^{\star}}\big]
\Big\rVert_2^2
+ \big(\mu_i - \mathrm{sg}[\mu_i^{\star}]\big)^2 + \big(\sigma_i - \mathrm{sg}[\sigma_i^{\star}]\big)^2 ,
\label{eq:shape}
\end{equation}
averaged over the $N$ tokens, the $L_s$ samples of each shape and the batch. The first term asks the codebook to describe the \textbf{shape} of the sub-series and the second asks the attribute head to describe its offset and scale, which is the split that makes the attributes interpretable. Matching the waveform after normalization is an important choice. If we were to compute the loss directly between the unnormalized reconstruction $\hat{s}_i = \bar{s}_i \sigma_i + \mu_i$ and the target segment $s_i$, the gradient of the shape matching objective would be scaled by the standard deviation $\sigma_i$. Consequently, whenever a token converges on a relatively flat region of the signal, the gradients would vanish, effectively slowing the learning process for that token.
\subsubsection{Disentanglement.} In the presented formulation, there is no mechanism to prevent all $N$ tokens from collapsing to represent the exact same segment of the input window. To ensure a diverse coverage of the signal, we introduce a spatial penalty to disentangle the tokens. We achieve this by embedding the temporal support of each token as a polar coordinate system:
\begin{equation}
v_i = a_i\,\big(\cos \pi t_i,\; \sin \pi t_i\big),
\qquad a_i = \frac{\log l_i}{\log l_{\min}},
\label{eq:polar}
\end{equation}
In this representation, the angular component encodes the token position in time, while the radial distance $a_i \in (0, 1]$ encodes \textbf{how long} the token's segment is. Specifically, the radius is maximized ($a_i = 1$) for the shortest admissible segment length ($l_i = l_{\min}$), and it shrinks toward the origin ($a_i \rightarrow 0$) as the segment length expands to span the entire window ($l_i \rightarrow 1$).

To encourage tokens to spread out and capture distinct features, we penalize pairs of tokens whose embeddings fall closer together than a predefined margin $\epsilon$:
\begin{equation}
\mathcal{L}_{\mathrm{dist}} = \frac{1}{N(N{-}1)}\sum_{i \neq j}
\big[\epsilon - \lVert v_i - v_j \rVert_2\big]_{+},
\label{eq:dist}
\end{equation}
where $[\,\cdot\,]_{+} = \max(0, \cdot)$ and $\epsilon{=}0.2$. The polar embedding is deliberate and it makes the penalty scale-aware, so two tokens may share a position provided they describe shapes of different length.

\subsection{Downstream adaptation}
\label{sec:adapt}
Task $t$ provides samples $(x_i, y_i)$ with $y_i \in \{1,\dots,C_t\}$ (classification), $y_i \in \mathbb{R}^{d_t}$ (regression), or both (multi-task). The frozen backbone is applied to one channel at a time and the channels are concatenated,
\begin{equation}
\phi(x) \;=\; \Big[\, g\big(f_\theta(r_1(x))\big); \;\cdots\;;
g\big(f_\theta(r_{M_t}(x))\big) \Big] \;\in\; \mathbb{R}^{M_t d_g},
\label{eq:phi}
\end{equation}
where $r_m(x)$ is the $m$-th channel produced by $\mathcal{R}_t$, $d_g$ is the output width of the descriptor $g$ ($d_g \in \{D,\, N(d_c{+}4),\, 4N,\, K\}$ depending on which descriptor is used), and $[\,\cdot\,; \,\cdot\,]$ denotes concatenation. Both $f_\theta$ and $g$ are shared across channels and tasks and only $M_t$ and the choice of $g$ change between tasks, which is why one univariate model serves 2-channel IQ, 3-channel IQA, 4-channel composites and 45-channel multi-anchor CIR stacks without an architectural change. A lightweight head $h_\psi$ is then trained by minimizing
\begin{multline}
\min_{\psi}\; \frac{1}{|\mathcal{D}_t|}\sum_{i}
\Big[\, \mathbf{1}_{\mathrm{cls}}\,\mathrm{CE}\big(h_\psi^{c}(\phi(x_i)), y^c_i\big)
+  \\ \mathbf{1}_{\mathrm{reg}}\,\ell_\delta\big(h_\psi^{r}(\phi(x_i)) - y^r_i\big) \Big],
\label{eq:downstream}
\end{multline}
where $\mathbf{1}_{\mathrm{cls}}$ and $\mathbf{1}_{\mathrm{reg}}$ indicate whether task $t$ has a classification and a regression part, $h_\psi^{c}$ and $h_\psi^{r}$ are the corresponding output heads, $y^c_i$ and $y^r_i$ the corresponding targets, $\mathrm{CE}$ is cross-entropy, and $\ell_\delta$ is the Huber loss with transition point $\delta$. Regression targets are $z$-scored with training-split statistics and reported back in physical units.

The multi-task prediction head yields $J$ distinct objective terms. For instance, when applied to the RadChar dataset, this results in $J=5$ separate components. Because these individual loss terms possess entirely different scales and units, their raw values are not directly comparable and cannot be trivially summed. To prevent the terms with the largest magnitudes from dominating the gradient updates, we dynamically balance the overall multi-task objective by combining them using learned homoscedastic uncertainty weights~\cite{kendall2018},
\begin{equation}
\mathcal{L}_{\mathrm{mt}} = \sum_{j=1}^{J}
\frac{1}{2 \zeta_j^{2}}\,\mathcal{L}_j + \log \zeta_j ,
\label{eq:kendall}
\end{equation}
where $\mathcal{L}_j$ is the $j$-th term of~\eqref{eq:downstream} and $\zeta_j > 0$ is a learned scalar, interpretable as that term's noise level, optimized jointly with $\psi$. A term the model cannot fit is automatically down-weighted by a large $\zeta_j$, and the $\log \zeta_j$ penalty is what prevents the trivial solution $\zeta_j \rightarrow \infty$.

Low-rank adaptation (LoRA) additionally adapts a low-rank correction to selected projections of the frozen backbone~\cite{hu2022lora}. For a weight matrix $W \in \mathbb{R}^{d_{\mathrm{out}} \times d_{\mathrm{in}}}$,
\begin{multline}
W' = W + \Delta W, \quad \Delta W = \tfrac{\alpha}{r} B A, \quad \\
B \in \mathbb{R}^{d_{\mathrm{out}} \times r},\;
A \in \mathbb{R}^{r \times d_{\mathrm{in}}},
\label{eq:lora}
\end{multline}
where $r \ll \min(d_{\mathrm{in}}, d_{\mathrm{out}})$ is the rank, $\alpha$ a fixed scaling that decouples the update magnitude from $r$, and only $A$ and $B$ are trained. Initializing $B{=}0$ makes $\Delta W {=} 0$ at step zero, so training starts exactly at the pretrained function; each adapted projection then adds $r(d_{\mathrm{in}}{+}d_{\mathrm{out}})$ trainable values while $W$ itself stays frozen.

\section{RFShape}
\label{sec:rfshape}
\subsection{Architecture: Learning from the Shapes}
RFShape uses a shape-codebook time-series backbone~\cite{wen2024vqshape}, depicted in Fig.~\ref{fig:arch}. Our architecture uses a Transformer encoder over patches of the normalized univariate series, then a cross-attention tokenizer summarizes the window into $N$ latent shape tokens and a codebook of $K$ codes. We used a Transformer and shape decoder only during pretraining. 
Each quantized token contains an interpretable attribute tuple consisting of four components: a relative starting position \( t \), a relative length \( l \) (both of which are normalized to the total series length), an offset \( \mu \), and a scale \( \sigma \). This structure reveals both the location of a shape within the data and its duration in the latent space, which is effective for tasks related to timing. The complete model has 9.48\,M parameters, of which 7.4\,M (the encoder and tokenizer of Fig.~\ref{fig:arch}(a)) are retained downstream.

\begin{table}[!t]
\caption{Complete configuration and training parameters.}
\label{tab:hyper}
\centering
\footnotesize
\setlength{\tabcolsep}{4pt}
\begin{tabularx}{\columnwidth}{@{}l X l@{}}
\toprule
 & Quantity & Value \\
\midrule
\multicolumn{3}{@{}l}{\textbf{Architecture}} \\
$T$          & model input length                & 512 \\
$L_p$ / $P$  & patch length / patches per window & 8 / 64 \\
$D$          & embedding dimension               & 256 \\
             & encoder layers / heads            & 4 / 4 \\
             & tokenizer layers / heads          & 4 / 4 \\
             & decoder layers / heads (pretrain only) & 2 / 4 \\
$N$          & shape tokens                      & 32 \\
$K$ / $d_c$  & codebook size / code dimension    & 1024 / 8 \\
$L_s$        & shape length                      & 128 \\
$l_{\min}$   & minimum token support             & 1/8 \\
             & parameters (total / retained)     & 9.48 / 7.4\,M \\
\midrule
\multicolumn{3}{@{}l}{\textbf{Pretraining}} \\
$\mathcal{D}_0$ & dataset & synthetic NTN IQ \\
             & input channels                    & IQA \\
$L$          & window before resampling          & 4096 \\
$\rho$       & masked-patch fraction             & 0.185 \\
$\lambda_1, \lambda_2, \lambda_3$ & quantization / shape / dist.\ weights & 1, 1, 0.2 \\
$\beta$      & codebook-update weight                 & 1.0 \\
$\eta$ / $\gamma$ & entropy weight / marginal weight & 0.01 / 1.0 \\
$\kappa$     & soft-assignment temperature       & 0.01 \\
$\epsilon$   & disentanglement margin            & 0.2 \\
             & floor on $\sigma_i^{\star}$ (added to the s.d.) & $10^{-8}$ \\
             & optimizer                         & AdamW \\
             & learning rate                     & $3.89\times10^{-4}$ \\
             & schedule                          & 500-step warmup, cosine \\
             & weight decay / grad.\ clip        & 0.01 / 1.0 \\
             & batch $\times$ accumulation       & $128 \times 4$ \\
             & epochs / precision / seed         & 30 / FP32 / 42 \\
\midrule
\multicolumn{3}{@{}l}{\textbf{Downstream}} \\
             & probe                             & linear or 1-layer MLP \\
             & MLP hidden width                  & 256 \\
             & epochs / learning rate            & 120 / $3\times10^{-3}$ \\
             & batch size                        & 256 \\
             & weight-decay grid                 & $\{0, 0.01, 0.1, 1\}$ \\
             & dropout grid                      & $\{0, 0.2, 0.5\}$ \\
             & model selection                   & validation split \\
$r$ / $\alpha$ & LoRA rank / scaling             & 16 / 4 \\
             & LoRA epochs / learning rate       & 25 / $5\times10^{-4}$ \\
\bottomrule
\end{tabularx}
\end{table}

A normalized channel $u \in \mathbb{R}^{T}$ is split into $P = T/L_p$ non-overlapping patches of length $L_p$, linearly embedded and given a positional code, then passed through the Transformer encoder to $E \in \mathbb{R}^{P \times D}$. A cross-attention tokenizer with $N$ learned query vectors $\mathcal{Q} \in \mathbb{R}^{N \times D}$ compresses that $P$-length view into a fixed budget of $N$ tokens, and an attribute head $\mathrm{Attr}$ splits each token into a code vector and four scalars,
\begin{multline}
H = \mathrm{CrossAttn}(\mathcal{Q}, E, E) \in \mathbb{R}^{N \times D}, \\
\;\;
(z_e^i, t_i, l_i, \mu_i, \sigma_i) = \mathrm{Attr}(H_i),
\label{eq:tokenizer}
\end{multline}
where $\mathrm{CrossAttn}(\cdot,\cdot,\cdot)$ takes queries, keys and values in that order, $H_i \in \mathbb{R}^{D}$ is the $i$-th row of $H$, and $z_e^i \in \mathbb{R}^{d_c}$ is the token's pre-quantization code vector. The support is parameterized so that it is always a valid interval,
\begin{equation}
t_i = \tilde{t}_i\,(1 - l_{\min}), \qquad
l_i = \tilde{l}_i\,(1 - t_i) + l_{\min},
\label{eq:support}
\end{equation}
where $\tilde{t}_i, \tilde{l}_i \in [0,1]$ are the raw head outputs, which a sigmoid confines to that interval, and $l_{\min}$ is the shortest admissible support. The same head emits $\mu_i$ unconstrained and $\sigma_i$ through a softplus, so the predicted scale is strictly positive. This gives $t_i \in [0, 1{-}l_{\min}]$ and $t_i + l_i \leq 1$ by construction, so no token can point outside the window. After quantization the token is decoded back to a waveform and placed,
\begin{equation}
\hat{s}_i = \bar{s}_i\,\sigma_i + \mu_i ,
\label{eq:shapedec}
\end{equation}
where $\bar{s}_i = \mathrm{Dec}_s(z_q^i)$ and $\mathrm{Dec}_s: \mathbb{R}^{d_c} \rightarrow \mathbb{R}^{L_s}$ is the shape decoder of~\eqref{eq:shape}, which produces a zero-mean unit-variance waveform that $\sigma_i$ and $\mu_i$ then place. A token therefore reads as a four-part answer including \textbf{which} shape ($k_i$), \textbf{where} it sits ($t_i$), \textbf{how long} it is ($l_i$), and at what \textbf{offset and scale} ($\mu_i, \sigma_i$). 

\begin{table*}[!tb]
\caption{Downstream tasks, datasets, and the per-task input encoding
(Sec.~\ref{sec:repops}).\label{tab:tasks}}
\centering
\footnotesize
\begin{tabular}{llllll}
\toprule
\# & Task & Dataset & Targets & Input encoding & Adaptation \\
\midrule
1 & Interference detection & NTN testbed pairs & 2-class & IQAP & linear probe \\
2 & Radar modulation & DeepRadar2022~\cite{clerico2023} & 23-class & RADAR & linear probe \\
3 & Radar characterisation & RadChar-Small~\cite{huang2023} & 5-class + 4 regr. & IQ & linear probe; BitFit / LoRA / tokenizer FT \\
4 & UWB TDoA positioning & IIoT lab CIR~\cite{coppens2025} & 2-D position regr. & multi-anchor CIR & MLP probe; LoRA \\
5 & UWB ranging-error corr. & IIoT \& University CIR~\cite{fontaine2024} & signed error regr. & CIR & linear probe \\
6 & UWB (N)LOS classification & IIoT \& University CIR~\cite{fontaine2024} & 2-class & CIR & linear probe \\
7 & Tx fingerprinting & WiSig ManySig~\cite{hanna2022} & 6-class, cross-RX & IQ & linear probe; LoRA \\
8 & Modulation classification & 37-class dataset & 37-class (6 fam.) & SPEC+ACF+QA+QI & linear probe; tokenizer FT \\
9 & WTR short range & short range~\cite{Lui2017, subray2023} & 4-class & FFT & linear probe \\
10 & WTR long range & long range~\cite{fontaine2020} & 4-class & IQA & linear probe \\
\bottomrule
\end{tabular}
\end{table*}

From a tokenized window we expose four latent descriptors $g$, each mapping the
token set to a fixed-length vector:
\begin{equation}
\begin{aligned}
g_{\mathrm{enc}}    &= \tfrac{1}{N}\textstyle\sum_i H_i
                     &&\in \mathbb{R}^{D},\\
g_{\mathrm{tok}}    &= \big[\,z_q^i, t_i, l_i, \mu_i, \sigma_i\,\big]_{i=1}^{N}
                     &&\in \mathbb{R}^{N(d_c + 4)},\\
g_{\mathrm{attr}}   &= \big[\,t_i, l_i, \mu_i, \sigma_i\,\big]_{i=1}^{N}
                     &&\in \mathbb{R}^{4N},\\
g_{\mathrm{hist}}   &= \Big[\tfrac{1}{N}\textstyle\sum_i
                        \mathbf{1}[k_i {=} k]\Big]_{k=1}^{K}
                     &&\in \mathbb{R}^{K}.
\end{aligned}
\label{eq:readouts}
\end{equation}

Here $H_i$ is the tokenizer output of~\eqref{eq:tokenizer}, $z_q^i$ and $k_i$ the quantized code and its index, and $\mathbf{1}[\cdot]$ the indicator, so $g_{\mathrm{hist}}$ is the codebook-usage histogram normalized to sum to one.     These four latent readouts are called \textbf{tokenizer}, \textbf{token}, \textbf{token-attr} and \textbf{histogram} in the experiments. The dimensions in~\eqref{eq:readouts} are per-channel descriptor widths, denoted by $d_g$. Following~\eqref{eq:phi}, descriptors from the $M_t$ input channels are concatenated, producing a task feature vector $\phi(x) \in \mathbb{R}^{M_t d_g}$. Thus, the prediction head receives $M_t d_g$ features, whereas $d_g$ always denotes the width for a single channel. For WiSig, $M_t{=}2$, $D{=}256$, $N{=}32$, $d_c{=}8$, and $K{=}1024$. The per-channel widths of the tokenizer, token, token-attribute, and histogram readouts are therefore $256$, $384$, $128$, and $1024$, respectively. After concatenation across the two channels, the corresponding prediction-head input dimensions are $512$, $768$, $256$, and $2048$. To summarize the latent representations:
\begin{itemize}
    \item $g_{\mathrm{enc}}$ is taken pre-quantization and therefore avoids information loss from the codebook bottleneck, at the cost of mean-pooling the $N$ tokens into a single summary.
    \item $g_{\mathrm{tok}}$ passes the shape identity through the $K$-code bottleneck but carries the four attributes $(t_i, l_i, \mu_i, \sigma_i)$ unquantized, so it keeps \textbf{which shape}, \textbf{where} and
    \textbf{how big}.
    \item $g_{\mathrm{attr}}$ discards the shape identity entirely and retains only the token attributes $(t_i, l_i, \mu_i, \sigma_i)$, that is \textbf{where} a shape sits and \textbf{how big} it is, which is why it survives on timing tasks and degrades on fine waveform discrimination.
    \item $g_{\mathrm{hist}}$ is permutation-invariant in $i$, so it retains \textbf{which shapes} occurred but not where, competitive exactly when the class is a global usage pattern.
\end{itemize}

\subsection{Pretraining}
RFShape is pretrained on \textbf{fully synthetic} IQ data generated in MATLAB, including NB-IoT uplink, 5G NTN uplink, and 5G terrestrial downlink waveforms, all passed through an NTN channel. NB-IoT is produced with the Long Term Evolution (LTE) and Satellite Communications toolboxes and the two 5G sets with the 5G and Communications toolboxes. Each waveform is then passed through a 3GPP NTN tapped-delay-line (TDL) channel (profiles TDL-A, TDL-C, and TDL-D) \cite{3gpp38901} for a satellite in low Earth orbit (LEO) at $700$\,km altitude considering NTN assumptions \cite{3gpp38811}, including free-space path loss, atmospheric and clutter loss, receiver noise figure, and a satellite Doppler shift of up to $47$\,kHz, pre-compensated on the uplink to a residual near $0.9$\,kHz. The generator sweeps a range of link geometries and radio configurations. The specified parameters encompass elevation angles between $10^\circ$ and $90^\circ$, carrier frequencies of $1.98$, $2.0$, and $2.1$\,GHz, and subcarrier spacings of $15$ and $30$\,kHz. Channel bandwidths are 180 kHz for NB-IoT and {10,15,20,25,30} MHz for both the 5G NTN uplink and the 5G terrestrial downlink, each generated at 15 and 30 kHz subcarrier spacing. The modulation families include phase-shift keying (PSK) and quadrature amplitude modulation (QAM). The process varies the modulation schemes (from binary PSK (BPSK) and quadrature PSK (QPSK) up to 256-QAM), terminal speeds ($0$--$30$\,m/s), receiver noise figures ($3$--$10$\,dB), and macro-environments (rural, suburban, and urban). Altogether, this comprehensive sweep produces $97{,}200$ distinct scenarios used in this study, yielding mean per-scenario SNR spanning a wide dynamic range from $-40$ to $+53$\,dB. A metadata filter then keeps the better-conditioned links (elevation $\geq 45^\circ$, noise figure $\leq 5$\,dB, atmospheric loss $\leq 1$\,dB, rural and suburban only), leaving $401{,}940$ windows, $1.65\times10^{9}$ complex samples in total, split $70/15/15$ and composed of $49\%$ 5G NTN uplink, $26\%$ NB-IoT and $25\%$ 5G terrestrial downlink. Windows of 4096 complex samples are presented to the encoder through the same representation operator used downstream~\eqref{eq:repop}, instantiated here as the IQA encoding $[r_{\mathrm{I}}, r_{\mathrm{Q}}, r_{\mathrm{A}}]$ with $M_t{=}3$ and resampled to the model length $T{=}512$. Table~\ref{tab:hyper} gives the complete configuration of the model and training processes.

\begin{figure}[!t]
\centering
\captionsetup[subfloat]{font=scriptsize}
\subfloat[WiSig transmitters]{\includegraphics[width=0.7\columnwidth]{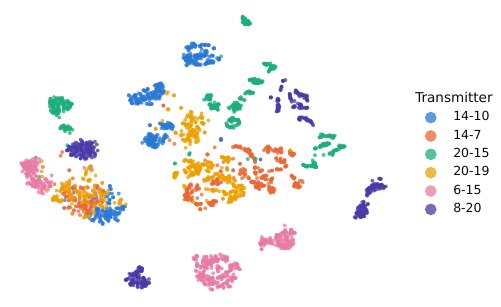}}
\hfil
\subfloat[DeepRadar modulations]{\includegraphics[width=0.7\columnwidth]{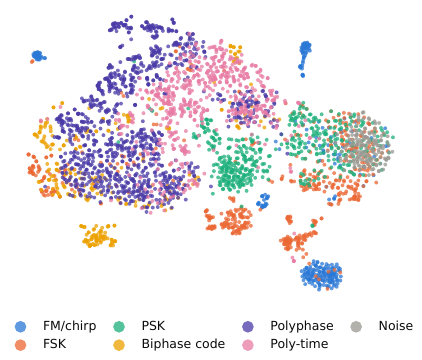}}
\hfil
\subfloat[UWB TDoA position]{\includegraphics[width=0.7\columnwidth]{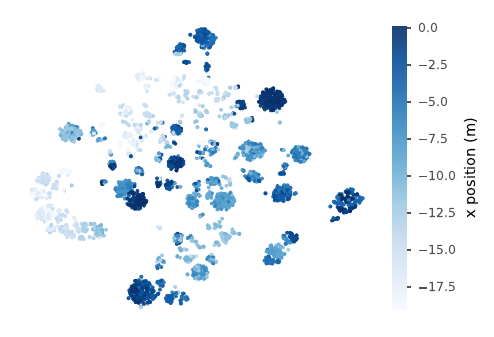}}
\caption{t-SNE of frozen RFShape features (tokenizer representation) on unseen
downstream data. (a) and (b) are coloured by class; (c) has no class label and
is coloured by the tag's measured $x$ coordinate.}
\label{fig:tsne}
\end{figure}

\subsection{Downstream Adaptation}
\subsubsection{Downstream tasks}
\label{sec:tasks}
Table~\ref{tab:tasks} lists the ten tasks, their datasets, and the input representation selected for each. Representation selection follows a simple principle of giving the frozen model a view in which the class-defining structure is stable. DeepRadar redraws its carrier per signal, so raw IQ of one class differs window to window; a carrier-invariant composite of log-envelope, carrier-removed instantaneous frequency, autocorrelation magnitude, and centered log-spectrum (the \textbf{RADAR} encoding) removes that nuisance. UWB tasks feed the CIR directly. Technology recognition at short range benefits from an FFT view. Adaptation strategies range from frozen linear/MLP probes over the latent representations, through feature fusion (e.g., concatenating DW1000 receiver diagnostics), to LoRA adapters on the backbone.
\subsubsection{Latent spaces}
Each panel projects the frozen $g_{\mathrm{enc}}$ descriptor of at most 4000 test-split windows. The descriptors were standardized, reduced to 50 components using principal component analysis (PCA), and then projected to two dimensions using t-distributed stochastic neighbor embedding (t-SNE) with perplexity 30, PCA initialization, and a fixed random seed. Fig.~\ref{fig:tsne} visualizes the frozen latent space on three tasks unseen for the model. WiSig transmitter fingerprints form compact clusters by transmitter identity even though every transmitter sends identical packet content, so only hardware impairment separates them. On DeepRadar, the 23 modulation classes exceed what any categorical palette can separate, so the projection is coloured by the six modulation families, including chirp, frequency-shift keying (FSK), phase-coded, and polyphase waveforms occupy distinct regions, with the noise class isolated. The third panel asks a different question, because UWB TDoA has no class to colour by. The same projection is coloured by the tag's measured $x$ coordinate. The embedding breaks into many compact clusters whose colour is consistent inside a cluster and varies smoothly between neighbouring ones, and the far end of the trajectory ($x \approx -18$\,m) occupies its own region of the map. The organization is local rather than one global gradient, but it is what makes the position regression of Sec.~\ref{sec:tdoa} possible.

\begin{figure}[!t]
\centering
\includegraphics[width=0.75\columnwidth]{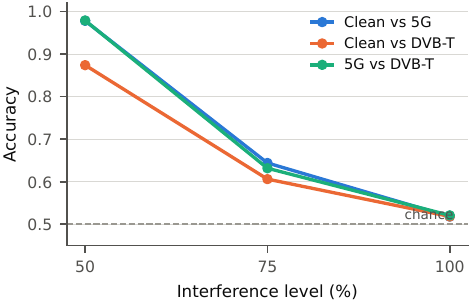}
\caption{Interference detection accuracy vs.\ interference level (frozen
linear probe, tokenizer representation).}
\label{fig:interf}
\end{figure}
\begin{figure}[!t]
\centering
\includegraphics[width=0.75\columnwidth]{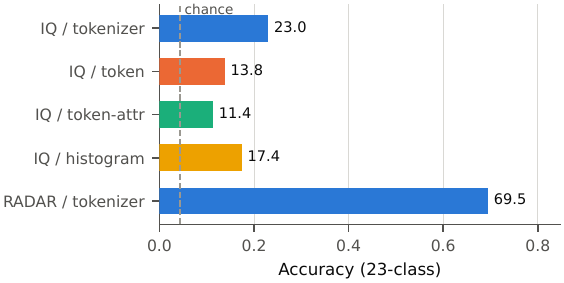}
\caption{DeepRadar 23-class accuracy. Comparison of raw-IQ views vs.\ the carrier-invariant RADAR encoding using the same frozen model.}
\label{fig:deepradar}
\end{figure}

\section{Experiments}
\label{sec:EXp}
\subsection{Setup}
All experiments use one pretrained checkpoint (dim 256, 4 encoder layers, 1024 codes). Probes are trained with AdamW, cosine decay, and grid-searched weight decay/dropout. Splits are leakage-controlled per task. For example, cross-receiver for WiSig, cross-environment for UWB error correction, cross-trajectory for TDoA, and stratified based on class and SNR for DeepRadar and RadChar. The used hardware is one NVIDIA RTX 5080 graphics processing unit (GPU) and a 24-thread central processing unit (CPU) \cite{gpulab}.
\subsection{Per-task results}
\subsubsection{Interference detection}
Binary discrimination between clean and interfered NTN captures (Clean vs.\ 5G, Clean vs.\ DVB-T, 5G vs.\ DVB-T) at three interferer powers, emulated with the CELEOS channel emulator. The CELEOS channel emulator~\cite{celeos} was configured for a LEO satellite at an altitude of 500~km, with varying elevation angles and atmospheric loss. The level index is an attenuation, including level 50, which is the strongest interferer, raising in-band power by 19.1 dB (5G) and 13.1 dB (DVB-T) over the clean capture, while at level 100 the interferer is attenuated to the point where the capture is indistinguishable from clean. Fig.~\ref{fig:interf} shows the frozen tokenizer probe results of 97.8\% (Clean/5G), 87.4\% (Clean/DVB-T) and 97.8\% (5G/DVB-T) at level 50, falling to chance at level 100, where the two classes are physically the same signal. The Clean/DVB-T combination presents the greatest challenge across all levels. This difficulty arises because the DVB-T interferer introduces 6 dB less in-band power compared to 5G. 

\subsubsection{DeepRadar (23-class radar modulation)}
\label{sec:deepradar}
On raw IQ, the frozen probe reaches 23.0\% accuracy over 23 classes. Because DeepRadar randomly changes the carrier frequency for every individual signal, the IQ data for any given signal category constantly fluctuates, creating an unpredictable pattern that is difficult to consistently classify.

The \textbf{RADAR} composite removes the nuisance before the encoder sees it. Each of its four maps is invariant, or nearly so, to that rotation. The
log-envelope $r_{\mathrm{LOGA}}$ and the autocorrelation magnitude
$r_{\mathrm{ACF}}$ discard the carrier exactly, since $R_{zz}[\tau] \mapsto e^{j\omega\tau}R_{zz}[\tau]$ leaves $\lvert R_{zz}\rvert$
unchanged, while $r_{\mathrm{CIF}}$ and $r_{\mathrm{SPECC}}$ estimate the offset by a circular mean and de-rotate before taking the instantaneous frequency and the spectrum. Measuring the mean relative $L_2$ change of each channel under carrier offsets of $\{0.10, -0.18, 0.23\}$ cycles per sample quantifies the gap: raw IQ drifts by $1.41$, whereas the composite drifts by $1.9\times10^{-4}$, four orders of magnitude less. The encoder is therefore shown the same series for the same class, whatever carrier the signal was drawn with.

Feeding to the same frozen model lifts the linear probe from 23.0\% to \textbf{69.5\%} against a 4.3\% chance level, a
$3.0\times$ improvement with no change to the weights (Fig.~\ref{fig:deepradar}). There are two observations about the general capacity. First, the parameter-free $k$-nearest-neighbor ($k$-NN) probe rises in steps, from 21.6\% to 59.2\%, showing that the gain is in the frozen features, not in a trained head. Second, every latent view improves together, and their ordering is preserved, which is what one expects if the input, not the descriptor, was the binding constraint.

\begin{figure*}[htbp]
\centering
\includegraphics[width=0.8\textwidth]{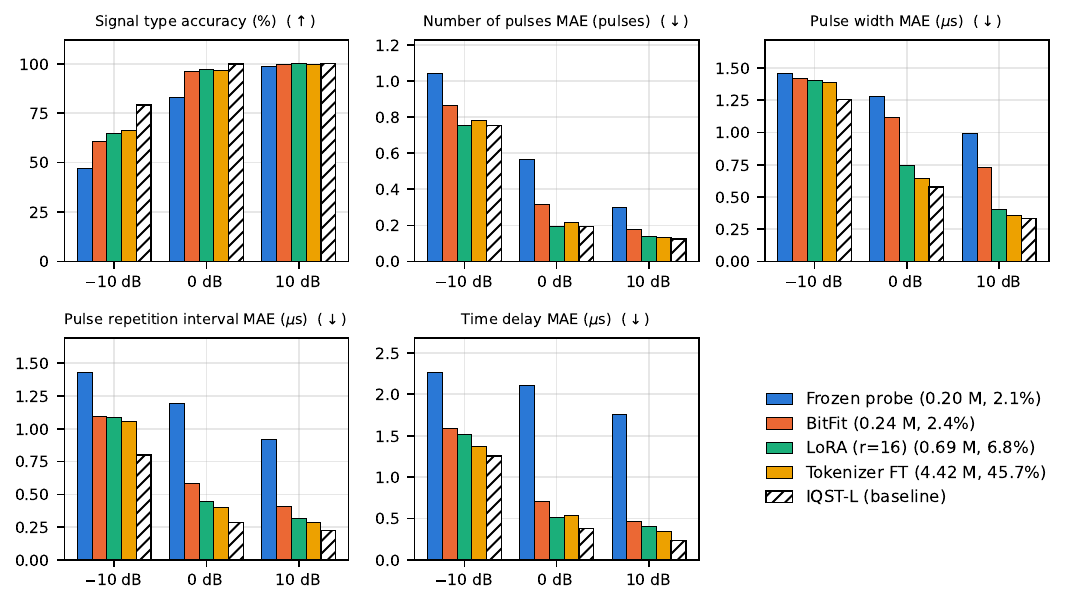}
\caption{RadChar adaptation against IQST-L~\cite{huang2023} at three SNRs. Each variant is adapted on 30\,k RadChar-Small frames, about $23\times$ fewer than the $\approx$700\,k RadChar-Baseline frames IQST-L is trained on.}
\label{fig:radchar_ft}
\end{figure*}
\subsubsection{RadChar (joint classification and timing regression)}
RadChar pairs one classification task (five signal types) with four timing regression tasks and it tests whether the shape tokens retain temporal localization. A frozen tokenizer probe reaches 74.2\% signal-type accuracy and regresses the four parameters to 0.669 pulses, 1.240\,$\mu$s, 1.192\,$\mu$s and 2.042\,$\mu$s mean absolute error (MAE) pooled over SNR. Consistent with the design of the latent space, the quantized token representation which carries the explicit $(t, l)$ attributes, is the better frozen feature for the two periodic targets (pulse repetition interval \textbf{0.956} vs.\ 1.192\,$\mu$s, time delay \textbf{1.305} vs.\ 2.042\,$\mu$s); On the other hand, the tokenizer embedding remains the stronger classifier.

Fig.~\ref{fig:radchar_ft} compares our adapted models against IQST-L, the strongest model of~\cite{huang2023}, which introduced RadChar and reports at $-10$, $0$ and $10$\,dB. The comparison is deliberately unequal because IQST-L trains end-to-end on 700\,k waveforms of RadChar-Baseline; On the other hand, RFShape adapts a radar-free pretrained encoder on 30\,k frames of RadChar-Small, $23\times$ less labelled data. IQST-L stays ahead at low SNR, 79.1\% against 66.3\% at $-10$\,dB and 99.8\% against 96.9\% at $0$\,dB, converging at $10$\,dB. Adaptation is worth most where the frozen probe is weakest, lifting accuracy from 47.1\% to 66.3\% at $-10$\,dB and from 82.9\% to 96.6\% at $0$\,dB, but adding about one point at $10$\,dB where the probe already reaches 98.8\%. Average of the accuracies over the SNRs shows a 6.6 to 8.3 point gain including BitFit 80.8\%, tokenizer fine-tuning 82.2\%, LoRA 82.5\%. The ordering is unstable, with tokenizer fine-tuning leading at $-10$\,dB on 66.3\% over LoRA at 64.6\% and BitFit at 60.9\%, and also roughly halving the regression errors at $0$ and $10$\,dB.

That residual shortage is mainly in classification, where the best variant falls 12.8 points behind on signal type at $-10$\,dB; On the other hand, it meets IQST-L's reported 0.752 pulses on pulse count and comes within $1.10\times$ and $1.11\times$ on time delay and pulse width; only the pulse repetition interval stays clearly behind, which no variant brings below $1.26\times$ at any SNR. Low SNR therefore corrupts which signal is present more than where its pulses lie. A shape token exposes its support $(t, l)$ directly, while quantization discards the waveform detail that separates the five types. 

\begin{figure}[!t]
\centering
\includegraphics[width=0.9\columnwidth]{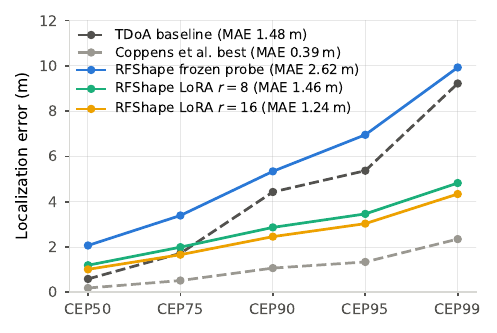}
\caption{UWB TDoA circular error probability (CEP), train \textbf{racks} $\rightarrow$ eval \textbf{racks2}; MAE in the legend. Dashed grey curves are reported in~\cite{coppens2025}. LoRA-adapted RFShape beats the uncorrected TDoA baseline at every percentile above the median.}
\label{fig:tdoa_cep}
\end{figure}
\subsubsection{UWB TDoA positioning} \label{sec:tdoa}
The model regresses 2-D position from the 300 sample CIRs of all anchors that received the transmission (6.2 on average). Adaptation is performed on the \textbf{racks} trajectory and evaluated on the unseen \textbf{racks2} trajectory (using the exact data splits, input representations, and evaluation metrics shown in the published transformer-based TDoA correction baseline~\cite{coppens2025}). Their transformer is trained directly on this UWB data, using all 15,505 labelled positions of the \textbf{racks} trajectory, whereas RFShape keeps a backbone pretrained on synthetic NTN IQ that had never seen a channel impulse response, and uses the UWB data only to fine-tune rank-16 adapters and a regression head. The shortfall is a filtering difference, as we additionally discard motion-capture (MOCAP) dropouts and superframes received by fewer than three anchors, rather than deliberate subsampling. Fig.~\ref{fig:tdoa_cep} summarizes the comparison. With LoRA rank-16 adapters (0.49\,M adapter parameters, 9.32\% of the backbone), RFShape reaches 1.24\,m MAE, beating the deployed uncorrected TDoA baseline (1.48\,m, $+$15.9\%) and has better results in the error tail. TDoA has the better median (0.59\,m vs.\ 1.01\,m). Geometric multilateration is precise under good link geometry and catastrophic without it, while the learned patterns are uniformly sensible.
\subsubsection{UWB ranging-error correction} 
\begin{figure}[!t]
\centering
\includegraphics[width=\columnwidth]{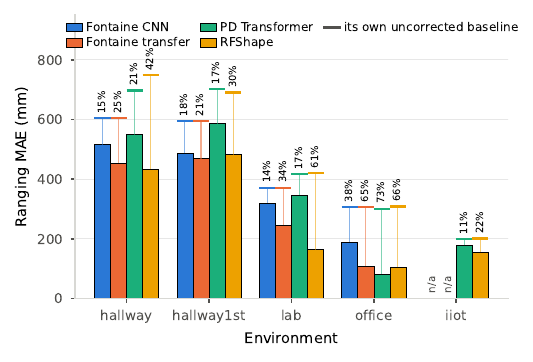}
\caption{MAE for UWB ranging across different environments, evaluated using a frozen linear probe on token representations. The baselines are the Fontaine convolutional neural network (CNN) and Fontaine transfer-learning models~\cite{fontaine2024} and the PD Transformer~\cite{cheraghinia2025}, quoted as reported. The colored caps indicate the uncorrected error for each setup, while the thin vertical lines show the error reduction achieved by each method.}
\label{fig:uwbenv}
\end{figure}

Using a 150-sample DW1000 CIR, the model performs regression-based error correction. The main metric is the uncorrected error that the model eliminates. We have evaluated across all environments of the dataset split ($n{=}11{,}418$); the frozen probe removes 39.0\% of the MAE, reducing the average error from 326\,mm to 199\,mm, while decreasing the 90th percentile (p90) error from 598\,mm to 392\,mm. Figure~\ref{fig:uwbenv} provides a per-environment analysis of these results, demonstrating error reductions from 22\% to 66\%. The three works do not share a common uncorrected error level; for example, in the same hallway it is $604$\,mm for~\cite{fontaine2024}, $697$\,mm for~\cite{cheraghinia2025} and $748$\,mm on our split. We could not reproduce a single common baseline, so Fig.~\ref{fig:uwbenv} reports every method against its own uncorrected value, and the comparison is in relative terms. Against the published per-environment corrections, RFShape shows better performance compared to PD Transformer~\cite{cheraghinia2025} in four of the five environments (it trails only in the office, 66\% against 73\%) and exceeds both Fontaine baselines~\cite{fontaine2024} in every environment they report. This is a considerable achievement, considering that \textbf{RFShape was never exposed to any CIR data during its pretraining}. The three methods also differ in what their labels buy. Both baselines are pretrained on CIRs drawn from this same dataset, supervised on a source environment for the Fontaine models~\cite{fontaine2024} and self-supervised on the office and iiot environments for the PD Transformer~\cite{cheraghinia2025}, and both then update the whole network for each target environment. RFShape's encoder sees no UWB data at any stage, so its labelled samples fit only the 1,153 parameters of a linear head on frozen features, using each environment's own training split.
\begin{figure}[!t]
\centering
\includegraphics[width=\columnwidth]{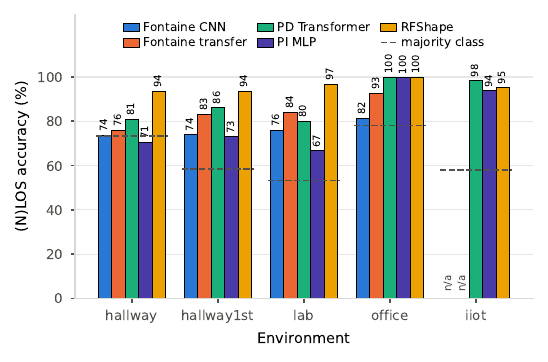}
\caption{UWB (N)LOS classification accuracy per environment (frozen linear probe). The baselines are the Fontaine CNN and Fontaine transfer-learning models~\cite{fontaine2024}, the PD Transformer~\cite{cheraghinia2025} and the PI MLP~\cite{cheraghinia2025lwfm}. The dashed cap is the majority-class rate, which differs per environment. The Fontaine baselines do not report the industrial hall, which was their source domain.}
\label{fig:uwbnlos}
\end{figure}
\subsubsection{UWB (N)LOS Classification}
\label{sec:uwbnlos}
This section evaluates a frozen linear model, adapted from RFShape, for UWB (N)LOS classification using a 150-tap DW1000 dataset. We compare our approach against several recent baselines, including those by \cite{fontaine2024}, \cite{cheraghinia2025}, and the lightweight PI MLP from \cite{cheraghinia2025lwfm}, as shown in Fig.~\ref{fig:uwbnlos}. 

Our results demonstrate that RFShape exhibits better generalization in unseen environments. It achieves accuracies of 93.7\% in the hallway, 93.7\% on the hallway1st, and 96.9\% in the lab, outperforming the strongest baseline by up to 12.7 percentage points. In contrast, the baseline models perform slightly better only in the specific environments they were pretrained on (the office and iiot). The PI MLP model, which has very few trainable parameters, in particular, struggles to generalize and shows a performance drop in unseen domains because of lack of scalability due to the size of the model.

Furthermore, raw accuracy is an incomplete metric due to significant class imbalances across environments. For instance, simply predicting the majority class yields a 78.1\% accuracy in the office, but only 53.3\% in the lab. When evaluated against this majority-class baseline (indicated by dashed caps in Fig.~\ref{fig:uwbnlos}), RFShape demonstrates substantial learning, lifting accuracy by 43.6 points in the lab. Conversely, models like PI MLP actually perform worse than the majority-class baseline in unseen environments like the hallway. Finally, we note that the quantized \textbf{token} representation provides the most effective features in four out of the five evaluated environments.
\subsubsection{WiSig Transmitter Fingerprinting}
\label{sec:wisig}
In this task, the model distinguishes between six Wi-Fi transmitters broadcasting exact preambles, meaning classification relies completely on hardware impairments. Using a cross-receiver split, a frozen linear probe applied to the quantized-token representation achieves 78.6\% accuracy, far exceeding the 16.7\% random chance baseline. Label-free evaluations (such as a $k$-NN probe with $k{=}5$ scoring 60.1\% and unsupervised $k$-means clustering) confirm that this discrimination is inherent to the representation. However, the features are linearly separable rather than naturally separated, given that the trained probe significantly outperforms the parameter-free $k$-NN approach.

Table~\ref{tab:wisigtune} explores strategies to improve performance while keeping the pretrained weights frozen. First, modifying the input encoding to include amplitude alongside raw IQ data ($[r_{\mathrm{I}}, r_{\mathrm{Q}}, r_{\mathrm{A}}]$) provides a 2.4-point accuracy boost. This simple approach outperforms the IQAP encoding (79.7\%), which appends the differential-phase channel $r_{\mathrm{DP}}$ because oscillator impairment is a phase effect. However, while additional channels benefit a trained classification head, they also disperse the embedding space. As a result, unsupervised clustering agreement decreases, with the normalized mutual information (NMI) dropping from 0.450 to 0.412.

Adapting the backbone yields the highest performance. Applying rank-16 LoRA adapters directly to raw IQ inputs reaches \textbf{83.9\%}, a 5.3-point gain. Notably, combining these approaches is not beneficial; using LoRA with the IQAP encoding scores 83.5\%, underperforming plain IQ and IQA with LoRA.
\begin{table}[!t]
\caption{WiSig cross-receiver accuracy under two adaptation methods.\label{tab:wisigtune}}
\centering
\footnotesize
\begin{tabular}{llcc}
\toprule
Input encoding & Adaptation & Trainable & Accuracy \\
\midrule
IQ          & frozen probe     & head only        & 78.6\% \\
IQAP & frozen probe     & head only        & 79.7\% \\
IQA         & frozen probe     & head only        &\textbf{81.0\%} \\
\midrule
IQAP & LoRA $r{=}16$    & 0.75\,M          & 83.5\% \\
IQ          & LoRA $r{=}16$    & 0.62\,M          & 83.9\% \\
IQA         & LoRA $r{=}16$    & 0.69\,M          & \textbf{85.7\%} \\
\bottomrule
\end{tabular}
\end{table}
\subsubsection{Modulation Classification}
\label{sec:mod37}
Our final evaluation presents the most challenging task which is a dataset from EDGX \cite{edgx_website}, that classifies 37 modulation schemes across six distinct families: on-off keying (OOK), amplitude-shift keying (ASK), pulse-amplitude modulation (PAM), PSK, QAM, and OFDM alongside a noise class. Specifically, classes within the same family differ only by their modulation order. For this task, the frozen model processes a specialized four-channel input consisting of a log-spectrum ($r_{\mathrm{SPEC}}$), log-lag autocorrelation ($r_{\mathrm{ACF}}$), and sorted quantile curves for amplitude ($r_{\mathrm{QA}}$) and real components ($r_{\mathrm{QI}}$).

Using a frozen linear probe on the tokenizer embedding, the model achieves a accuracy of $96.4 \pm 0.2$\% at identifying the broader modulation family, but only $45.4 \pm 0.8$\% for the exact fine class. Unfreezing the tokenizer improves fine-class accuracy to $50.0 \pm 0.9$\%, while family-level accuracy only slightly increases to $96.9 \pm 0.2$\%. This shows that \textbf{identifying the modulation family is effectively solved by the frozen features, but determining the specific modulation order requires further model adaptation.}

\begin{figure}[!t]
\centering
\includegraphics[width=\columnwidth]{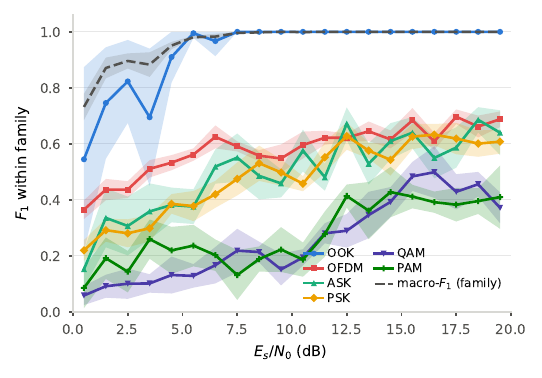}
\caption{Modulation order recovery against SNR for the fine-tuned model, averaged over ten seeds. Each family curve is the mean $F_1$ over the \textbf{fine classes within} that family. The dashed line is the family-level macro-$F_1$, computed after collapsing predictions and labels onto the seven families. OOK has a single class,
so for it the two coincide.}
\label{fig:modsnr}
\end{figure}

Figure~\ref{fig:modsnr} resolves that split against $E_s/N_0$. The two kinds of curve answer different questions. The dashed curve is the family-level macro-$F_1$, obtained by collapsing both the prediction and the label from the 37 fine classes onto the seven families and averaging the per-family $F_1$ without weighting, so it asks only whether the family was identified. Within-family difficulty depends on what separates the orders. OOK is the binary case of amplitude-shift keying, and the dataset keeps it as a class of its own. The remaining families span BPSK, QPSK and 8- to 64-PSK, 4- to 64-ASK, 4- to 64-PAM, 16- to 1024-QAM, and twelve OFDM configurations of 64 to 2048 subcarriers, where the within-family variable is the subcarrier count rather than a constellation order. The $F_1$ scores for OFDM and ASK reach approximately 0.70 and PSK 0.63, whereas QAM and PAM peak much lower (at 0.50 and 0.43, respectively) and fail to improve further up to 20\,dB. This plateau highlights an inherent challenge of the task rather than a model limitation. Higher order QAM and PAM constellations possess densely packed amplitude levels, making them fundamentally difficult to separate even under ideal conditions.
\begin{figure}[!t]
\centering
\includegraphics[width=\columnwidth]{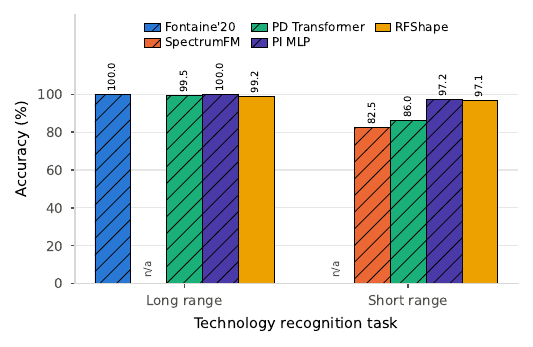}
\caption{Wireless technology recognition compared with prior work. The baselines are the task-specific model of Fontaine \emph{et al.}~\cite{fontaine2020}, SpectrumFM~\cite{zhou2026spectrumfm}, the PD Transformer~\cite{cheraghinia2025} and the PI MLP~\cite{cheraghinia2025lwfm}. RFShape has never observed these technologies, receivers, or datasets during pretraining.}
\label{fig:wtr}
\end{figure}
\subsubsection{Wireless technology recognition}
\label{sec:wtr}
The WTR bundle provides a test for transfer learning, allowing us to compare our approach with prior foundation models on standardized tasks. The target technologies (four long-range and four short-range), the capture sites, and the spectral view were entirely absent from our pretraining corpus. On the other hand, all baseline models presented in Fig.~\ref{fig:wtr} were trained on this dataset.

For long-range technologies, classification performance saturated across all methods. Our codebook histogram approach achieves 99.2\% accuracy performing comparably to the PD Transformer~\cite{cheraghinia2025} (99.53\%) and both the PI MLP~\cite{cheraghinia2025lwfm} and task-specific models~\cite{fontaine2020} (100\%). This demonstrates that a global codebook-usage vector effectively captures key identifying features, such as occupied bandwidth and burst structure, without requiring detailed waveform analysis (Sec.~\ref{sec:abl-repr}).

For short-range technologies, the choice of input representation is the determining factor. Using the same frozen encoder, time-domain inputs yield 70.8\% accuracy with naive resampling and 79.8\% with anti-aliasing. Switching to a spectral representation boosts accuracy to 97.1\% while requiring a tenth of the labelled data, 700 windows against 7032 for the time-domain view of the same captures. The adaptation cost is correspondingly small, since that 97.1\% comes from training a 3,076-parameter linear head on a frozen backbone that has never seen this corpus, whereas every baseline in Fig.~\ref{fig:wtr} is pretrained or fine-tuned on it. On that footing RFShape lands within 0.16 points of the best in-domain model (PI MLP, 97.25\%) and substantially ahead of models tied to a fixed input representation, such as the PD Transformer (86.03\%) and SpectrumFM~\cite{zhou2026spectrumfm} ($\approx$82.5\%).

These results show the main contribution of the paper that the classes are already separable within the frozen latent space if the signal is presented in an appropriate view. Exploring different input representations is computationally more efficient than pretraining a completely new model.
\begin{figure}[!t]
\centering
\includegraphics[width=\columnwidth]{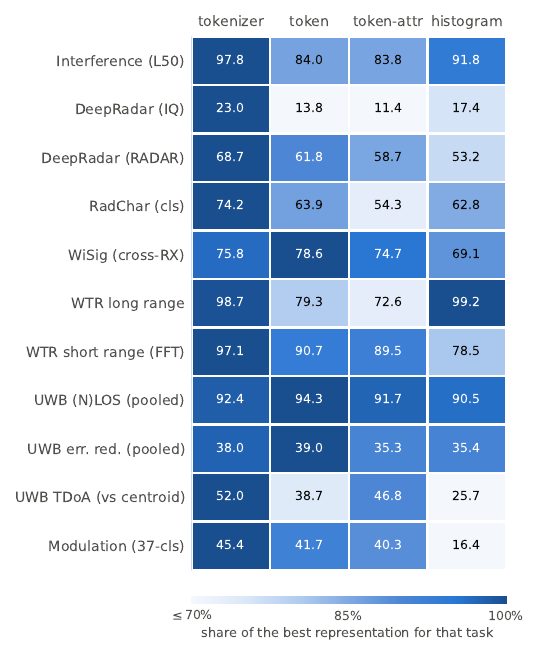}
\caption{Latent-representation ablation across all ten tasks. Cells are per-task metrics in \%: accuracy, except the two UWB rows, which are the reduction in ranging MAE and the improvement over a centroid-position baseline.}
\label{fig:heatmap}
\end{figure}

\subsection{Ablation studies}
\subsubsection{Latent readout}
\label{sec:abl-repr}
Fig.~\ref{fig:heatmap} sweeps the four latent readouts across eleven settings: the ten tasks of Table~\ref{tab:tasks}, with DeepRadar appearing twice because it is reported under both input encodings. No single latent readout wins everywhere.

The pre-quantization \textbf{tokenizer} embedding takes seven of the eleven rows by margins of 3 to 10 points. The discrete \textbf{token} representation takes the remaining three, by 1.0 to 2.8 points including transmitter fingerprinting (78.6\% vs.\ 75.8\%), UWB (N)LOS classification (94.3\% vs.\ 92.4\%) and UWB ranging-error reduction (39.0\% vs.\ 38.0\% of the MAE removed). This difference is not random but follows from what quantization does. Replacing $z_e^i$ by its nearest code discards the variation within a codebook cell outright rather than attenuating it, since the residual never reaches the head, while the attributes $(t_i, l_i, \mu_i, \sigma_i)$ pass through unquantized. Where that discarded variation carries the class, as on tasks that rely on intricate waveform shapes, accuracy falls. The two UWB rows ask for a property of the propagation channel rather than of one particular multipath realization, and the cross-receiver fingerprinting split asks for transmitter identity while the receiver itself changes between training and test (Sec.~\ref{sec:wisig}).

The \textbf{token-attr} representation is the most compact of the four, 128 values per channel against 384 for the full token, because it keeps the four attributes and drops the code. It is top on no task, but the cost of dropping the code depends entirely on whether the identity of a shape carries the class. That cost is largest on RadChar signal-type classification, 9.6 points below the full token, where the five classes differ by the shape of the pulse itself, and on long-range technology recognition, 6.6 points, where the class is in effect which codes the window uses, as the histogram result for the same task confirms. Token-attr costs only 0.2 points on interference detection and gains 8.0 points on TDoA position regression, the only one of the eleven settings in which discarding the code helps at all. Knowing when a component occurs and how long it lasts is therefore worth more than knowing its exact shape on exactly those tasks whose label is a location in time or space.

Furthermore, the \textbf{histogram} representation achieves the best result only in long-range technology recognition (99.2\%) and performs the poorest on half of the evaluated tasks. This confirms that global codebook-usage statistics are primarily effective when class distinctions rely on broad spectral occupancy patterns rather than fine signal details.
\subsubsection{Pretraining channel set}
\label{sec:abl-channels}
The pretraining input is itself a design choice. We pretrained two identical models, one on IQA $(I, Q, |z|)$ and one on IQAP, which appends the differential phase, and ran both through the identical 19-run downstream suite. Excluding the two alternative short-range encodings, which duplicate a task already counted, 17 measurements are directly comparable: \textbf{IQA wins 10 and IQAP 7} (Table~\ref{tab:channels} lists the eight headline tasks). The IQAP wins are concentrated in the ranging-error cells, five of the seven, but they are not confined to them: IQAP also takes transmitter fingerprinting (79.52\% vs.\ 78.58\%) and the frozen TDoA probe (2.43\,m vs.\ 2.62\,m). Adding an explicit phase channel to pretraining therefore costs downstream transfer on average rather than uniformly, and all results in this paper use the IQA checkpoint.
\begin{table}[!t]
\caption{Pretraining channel-set ablation, identical
architecture and schedule, differing only in the channels seen during
pretraining.\label{tab:channels}}
\centering
\footnotesize
\setlength{\tabcolsep}{4pt}
\begin{tabular}{ p{4.8cm} >{\centering\arraybackslash}p{0.6cm} >{\centering\arraybackslash}p{0.6cm} >{\centering\arraybackslash}p{1.6cm} }
\toprule
Task & IQAP & IQA & $\Delta$ (IQA$-$IQAP) \\
\midrule
WTR long-range        & 98.94 & \textbf{99.20} & $+$0.26 \\
WTR short-range (FFT) & 95.93 & \textbf{97.09} & $+$1.16 \\
UWB (N)LOS          & 89.91 & \textbf{91.44} & $+$1.53 \\
DeepRadar (RADAR)     & 67.84 & \textbf{69.51} & $+$1.67 \\
RadChar (class)       & 72.47 & \textbf{74.20} & $+$1.73 \\
WiSig (cross-RX)      & \textbf{79.52} & 78.58 & $-$0.94 \\
UWB err. red.\ (pooled, \% MAE reduction) & \textbf{39.74} & 39.02 & $-$0.72 \\
UWB TDoA MAE (m, LoRA $r{=}16$; lower better) & 1.26 & \textbf{1.24} & $+$0.02 \\
\bottomrule
\end{tabular}
\end{table}
\subsection{Deployment}
Table~\ref{tab:hw} reports measured inference characteristics of the tokenize path, the only path used downstream. What is actually deployed is the encoder and tokenizer of Fig.~\ref{fig:arch}(a), 7.40\,M of the 9.48\,M pretrained parameters, or \textbf{29.6\,MB} of FP32 weights; the decoder and shape heads, which exist only to compute the reconstruction term during pretraining, are discarded. The 114\,MB training checkpoint is roughly three times the weight size because it also stores the two Adam moments. The measured path is the tokenize path only. One univariate channel entering the encoder and leaving the tokenizer as the $N$ shape tokens of~\eqref{eq:tokenizer}. The representation operator $\mathcal{R}_t$ of~\eqref{eq:repop} is parameter-free and costs 0.16 to 0.21\,ms on the CPU for the encodings used here, and the task head costs 0.008\,ms. What makes the model suitable for operation close to the radio is less the latency than four properties of this path.

First, the per-task cost is negligible. One copy of the 7.40\,M parameter encoder is shared by every task, and a task is added by training a head of 1{,}153 parameters for ranging-error correction or 3{,}076 for short-range technology recognition, so a receiver that runs several of these functions stores one 29.6\,MB encoder and a few kilobytes per function instead of one network per function. Second, the number of input channels is not an architectural parameter, so the same deployed weights serve a two-channel IQ front end and the 45-channel multi-anchor stack without re-export; on the GPU the extra channels are nearly free, 9.4\,ms at 45 channels against 8.9\,ms at one, because they fold into the batch dimension. Third, retargeting the model to a new signal or view changes only $\mathcal{R}_t$, a deterministic transform with no weights and no training, so a deployed receiver can take on a new task without a new model being trained, shipped or stored. Fourth, the resident cost is small with 52\,MB of peak memory at batch 1 and under 1\,GB at batch 256, so inference needs no training framework on the device.
\begin{table}[!t]
\caption{Measured deployment characteristics (tokenize path, $T{=}512$; median over 50 runs). \label{tab:hw}}
\centering
\footnotesize
\begin{tabular}{lc}
\toprule
 & GPU (RTX 5080) \\
\midrule
Latency, batch 1 & 9.3\,ms \\
Throughput, batch 64 & 6{,}297 win/s \\
Throughput, batch 256 & 8{,}791 win/s \\
Peak memory, batch 1 / 256 & 52\,MB / 964\,MB \\
\midrule
Deployed parameters / weights & 7.40\,M / 29.6\,MB (FP32) \\
Per-task head, linear probe & 1.2--3.1\,k parameters \\
Pretrained total / checkpoint & 9.48\,M / 114\,MB \\
\bottomrule
\end{tabular}
\end{table}

\section{Conclusion}
This work presented RFShape, a wireless foundation model that extends the reuse of a shared pretrained encoder through deterministic input representation operators. Pretrained exclusively on synthetic IQ waveforms, without real captures, task labels, or CIR data, RFShape was evaluated on ten tasks across nine downstream datasets spanning classification, ranging-error regression, and two-dimensional positioning. The results demonstrate transfer to heterogeneous signal representations and to CIR inputs absent from pretraining, without requiring multimodal pretraining.

The experiments highlight the importance of input representation and latent readout selection. With the encoder frozen, changing the input representation increased radar classification accuracy from 23.0\% to 69.5\% and short-range technology recognition accuracy from 70.8\% to 97.1\%. Transfer to UWB tasks further demonstrated that features learned from synthetic IQ can support CIR-based classification and regression. However, no single representation or readout was optimal across all tasks, and additional adaptation improved performance in several settings. Representation selection and model adaptation should therefore be viewed as complementary rather than interchangeable. 

Overall, these findings show that the downstream applicability of a wireless foundation model need not be restricted to the representations or modalities included during pretraining. Domain-informed transformations can extend the utility of learned features while retaining a shared backbone.


\begin{thebibliography}{99}

\bibitem{cheraghinia2026wpfm}
M.~Cheraghinia \emph{et al.}, ``Wireless physical-layer foundation models:
Architectures, learning paradigms, applications, and deployment,''
arXiv:2608.20486, 2026.

\bibitem{jiang2026largeAI}
F.~Jiang \emph{et al.}, ``A Comprehensive Survey of Large AI Models for Future Communications: Foundations, Applications, and Challenges,'' in \emph{IEEE Communications Surveys \& Tutorials}, vol. 28, pp. 4731-4764, 2026, doi: 10.1109/COMST.2026.3660844.

\bibitem{aboulfotouh2024msm}
A.~Aboulfotouh, A.~Eshaghbeigi, D.~Karslidis, and H.~Abou-Zeid,
``Self-supervised radio pre-training: Toward foundational models for
spectrogram learning,'' in \emph{Proc. IEEE Global Commun. Conf. (GLOBECOM)},
2024, pp.~1--6, doi: 10.1109/GLOBECOM52923.2024.10901287.

\bibitem{aboulfotouh2025wavesfm}
A.~Aboulfotouh, E.~Mohammed, and H.~Abou-Zeid, ``6G WavesFM: A foundation
model for sensing, communication, and localization,'' \emph{IEEE Open J.
Commun. Soc.}, vol.~6, pp.~6792--6807, 2025, doi: 10.1109/OJCOMS.2025.3600616.

\bibitem{kim2026lwmspectro}
N.~Kim, S.~Alikhani, and A.~Alkhateeb, ``LWM-Spectro: A foundation model for
wireless baseband signal spectrograms,'' in \emph{Proc. IEEE Int. Conf.
Commun. (ICC)}, Glasgow, United Kingdom, 2026, pp. 1-6, doi: 10.1109/ICC59461.2026.11587404.

\bibitem{chen2025acsf}
S.~Chen, Z.~Feng, S.~Yang, Y.~Ma, J.~Liu, and Z.~Qi, ``A generative
self-supervised framework for cognitive radio leveraging time-frequency
features and attention-based fusion,'' \emph{IEEE Trans. Wireless Commun.},
2025, doi: 10.1109/TWC.2024.3513980


\bibitem{zhou2026spectrumfm}
F.~Zhou \emph{et al.}, ``SpectrumFM: A foundation model for intelligent
spectrum management,'' \emph{IEEE J. Sel. Areas Commun.}, vol.~44,
pp.~4471--4488, 2026, doi: 10.1109/JSAC.2025.3644783.

\bibitem{emind2026}
L.~Luo \emph{et al.}, ``EMind: A foundation model for multi-task
electromagnetic signals understanding,'' arXiv:2508.18785, 2025.

\bibitem{mashaal2025iqfm}
O.~Mashaal and H.~Abou-Zeid, ``IQFM---a wireless foundational model for I/Q
streams in AI-native 6G,'' arXiv:2506.06718, 2025.

\bibitem{chu2026wjepa}
V.~Chu, O.~Mashaal, and H.~Abou-Zeid, ``WirelessJEPA: A multi-antenna
foundation model using spatio-temporal wireless latent predictions,''
arXiv:2601.20190, 2026.

\bibitem{mashaal2026earlyexit}
O.~Mashaal and H.~Abou-Zeid, ``Fast wireless foundation models with
early-exits,'' \emph{IEEE Commun. Lett.}, vol.~30, pp.~2605--2609, 2026, doi: 10.1109/LCOMM.2026.3710677


\bibitem{aboulfotouh2026multimodal}
A.~Aboulfotouh and H.~Abou-Zeid, ``Multimodal wireless foundation models,'' in
\emph{Proc. IEEE Int. Conf. Commun. (ICC)}, 2026, doi:
10.1109/ICC59461.2026.11587642.

\bibitem{cheraghinia2025}
M.~Cheraghinia, E.~De Poorter, J.~Fontaine, M.~Debbah, and A.~Shahid, ``A
foundation model for wireless technology recognition and localization tasks,''
\emph{IEEE Open J. Commun. Soc.}, vol.~6, pp.~9879--9894, 2025, doi:
10.1109/OJCOMS.2025.3636436.

\bibitem{cheraghinia2025lwfm}
M.~Cheraghinia, E.~De Poorter, J.~Fontaine, K.~S. Kim, M.~Debbah, and
A.~Shahid, ``Lightweight foundation model for wireless time series downstream
tasks on edge devices,'' 2025 IEEE Globecom Workshops (GC Wkshps), Taipei, Taiwan, 2025, pp. 104-109, doi: 10.1109/GCWkshps68340.2025.11591243.


\bibitem{chen2025radiollm}
S.~Chen, Y.~Zu, Z.~Feng, S.~Yang, and M.~Li, ``RadioLLM: Introducing large
language model into cognitive radio via hybrid prompt and token
reprogrammings,'' arXiv:2501.17888, 2025.

\bibitem{wen2026rflm}
Z.~Wen, C.~Yang, H.~Zhang, Y.~Wang, H.~Shao, and S.~Wang, ``RFLM: An RF
language model for instruction-conditioned RF processing,'' \emph{IEEE Trans.
Mobile Comput.}, 2026, doi: 10.1109/TMC.2026.3715170.

\bibitem{zou2026rfgpt}
H.~Zou \emph{et al.}, ``RF-GPT: Teaching AI to see the wireless world,''
arXiv:2602.14833, 2026.

\bibitem{zou2026seeing}
H.~Zou \emph{et al.}, ``Seeing radio: From zero RF priors to explainable
modulation recognition with vision language models,'' in \emph{Proc. IEEE Int.
Conf. Commun. Workshops (ICC Workshops)}, 2026, doi:
10.1109/ICCWorkshops63917.2026.11586562.

\bibitem{han2026pred}
Z.~Han \emph{et al.}, ``PReD: An LLM-based foundation multimodal model for
electromagnetic perception, recognition, and decision,'' arXiv:2603.28183,
2026.

\bibitem{chi2024rfdiffusion}
G.~Chi \emph{et al.}, ``RF-Diffusion: Radio signal generation via
time-frequency diffusion,'' in \emph{Proc. 30th Annu. Int. Conf. Mobile Comput. Netw. (MobiCom)}, Washington D.C., USA, 2024, pp. 77–92, doi: 10.1145/3636534.3649348.


\bibitem{navidan2026oran}
H.~Navidan \emph{et al.}, ``Toward autonomous O-RAN: A multi-scale agentic AI
framework for real-time network control and management,'' \emph{IEEE Netw.},
2026, doi: 10.1109/MNET.2026.3701847.

\bibitem{huang2023}
Z.~Huang, A.~Pemasiri, S.~Denman, C.~Fookes, and T.~Martin, ``Multi-task
learning for radar signal characterisation,'' in \emph{Proc. IEEE Int. Conf.
Acoust., Speech, Signal Process. Workshops (ICASSPW)}, 2023, pp.~1--5, doi:
10.1109/ICASSPW59220.2023.10193318.

\bibitem{coppens2025}
D.~Coppens, A.~Shahid, and E.~De Poorter, ``UWB TDoA error correction using
transformers: Patching and positional encoding strategies,''
arXiv:2507.03523, 2025.

\bibitem{fontaine2024}
J.~Fontaine, M.~Ridolfi, B.~Van Herbruggen, A.~Shahid, and E.~De Poorter,
``Transfer learning for UWB error correction and (N)LOS classification in
multiple environments,'' \emph{IEEE Internet Things J.}, vol.~11, no.~3,
pp.~4085--4101, Feb. 2024, doi: 10.1109/JIOT.2023.3299319.

\bibitem{clerico2023}
V.~Clerico, J.~Gonz\'alez-L\'opez, G.~Agam, and J.~Grajal, ``LSTM framework
for classification of radar and communications signals,'' in \emph{Proc. IEEE
Radar Conf. (RadarConf23)}, San Antonio, TX, USA, 2023, pp.~1--6, doi:
10.1109/RadarConf2351548.2023.10149618.

\bibitem{hanna2022}
S.~Hanna, S.~Karunaratne, and D.~Cabric, ``WiSig: A large-scale WiFi signal
dataset for receiver and channel agnostic RF fingerprinting,'' \emph{IEEE
Access}, vol.~10, pp.~22808--22818, 2022, doi: 10.1109/ACCESS.2022.3154790.

\bibitem{kendall2018}
A.~Kendall, Y.~Gal, and R.~Cipolla, ``Multi-task learning using uncertainty to
weigh losses for scene geometry and semantics,'' in \emph{Proc. IEEE/CVF Conf.
Comput. Vis. Pattern Recognit. (CVPR)}, 2018, pp.~7482--7491, doi:
10.1109/CVPR.2018.00781.

\bibitem{hu2022lora}
E.~J. Hu \emph{et al.}, ``LoRA: Low-rank adaptation of large language
models,'' in \emph{Proc. Int. Conf. Learn. Represent. (ICLR)}, 2022.

\bibitem{wen2024vqshape}
Y.~Wen, T.~Ma, T.-W. Weng, L.~M. Nguyen, and A.~A. Julius, ``Abstracted shapes
as tokens---a generalizable and interpretable model for time-series
classification,'' in \emph{Proc. Adv. Neural Inf. Process. Syst. (NeurIPS)},
vol.~37, 2024.

\bibitem{edgx_website}
EDGX, ``EDGX home,'' 2026. [Online]. Available: \url{https://www.edgx.space/}.
[Accessed: Sep.~29, 2026].

\bibitem{Lui2017}
W.~Liu, M.~Kulin, T.~Kazaz, A.~Shahid, I.~Moerman, and E.~De Poorter,
``Wireless Technology Recognition Based on RSSI Distribution at Sub-Nyquist Sampling Rate for Constrained Devices,''
\emph{Sensors}, vol.~17, p.~2081, 2017, doi: 10.3390/s17092081.

\bibitem{subray2023}
S.~Subray, ``Real-world wireless communication dataset: IEEE 802.11ax, LTE, and 5G-NR signals,'' 2023. [Online]. Available: \url{https://dx.doi.org/10.21227/b82j-sy57}

\bibitem{fontaine2020}
J.~Fontaine, A.~Shahid, R.~Elsas, A.~Seferagic, I.~Moerman, and E.~De Poorter,
``Multi-band sub-GHz technology recognition on NVIDIA's Jetson Nano,'' in
\emph{Proc. IEEE 92nd Veh. Technol. Conf. (VTC2020-Fall)}, 2020, pp.~1--7, doi: 10.1109/VTC2020-Fall49728.2020.9348566.

\bibitem{3gpp38811}
3GPP, ``Study on New Radio (NR) to support non-terrestrial networks,''
3rd Generation Partnership Project, Technical Report TR 38.811,
V15.4.0, Rel.-15, Sep.\ 2020.

\bibitem{3gpp38901}
3GPP, ``Study on channel model for frequencies from 0.5 to 100 GHz,''
3rd Generation Partnership Project, Technical Report TR 38.901,
V17.0.0, Rel.-17, Mar.\ 2022.

\bibitem{gpulab}
IDLab, ``GPULab,'' Ghent University - imec. [Online]. Available: \url{https://idlab.ugent.be/resources/gpulab}. Accessed: Oct. 15, 2026.

\bibitem{celeos} LASTING Software, ``CELEOS,'' [Online]. Available: \url{https://www.lasting.space/celeos}. Accessed: Oct. 27, 2026.

\end{thebibliography}
\end{document}